\documentclass[a4paper,fleqn]{cas-dc}

\usepackage[authoryear]{natbib}

\def\tsc#1{\csdef{#1}{\textsc{\lowercase{#1}}\xspace}}
\tsc{WGM}
\tsc{QE}

\begin{document}
\let\WriteBookmarks\relax
\def\floatpagepagefraction{1}
\def\textpagefraction{.001}

\shorttitle{Frequency-enhanced Vision Transformer in Medical Image Segmentation} 

\shortauthors{Jin Yang}

\title [mode = title]{FEFormer: Frequency-enhanced Vision Transformer for Generic Knowledge Extraction and Adaptive Feature Fusion in Volumetric Medical Image Segmentation}  



%

\author[1]{Jin Yang}

\ead{yang.jin@wustl.edu}
\cortext[1]{Corresponding author}
\cormark[1]

\author[2]{Xiaobing Yu}

\author[2]{Peijie Qiu}


\affiliation[1]{organization={Department of Radiation Oncology},
            addressline={Icahn School of Medicine at Mount Sinai}, 
            city={New York},
            postcode={10029}, 
            state={NY},
            country={USA}}
            
\affiliation[2]{organization={Mallinckrodt Institute of Radiology},
            addressline={Washington University School of Medicine in St. Louis}, 
            city={St. Louis},
            postcode={63110}, 
            state={MO},
            country={USA}}


\begin{abstract}
Accurate segmentation of organs and lesions in medical images is essential for many clinical applications, and Vision Transformers (ViTs) have shown impressive segmentation performance. However, they face several key challenges in module and architecture designs. Specifically, self-attention is challenged from capturing fine-grained local features to understand detailed anatomical structures with large variations, and standard MLP modules lack explicit mechanisms to preserve detailed spatial information. Moreover, conventional encoder–decoder architectures rely on naive feature fusion strategies (e.g., concatenation), and they lack explicit mechanisms to align features and handle large semantic discrepancies for robust fusion. These architectures also downsample input images to lower dimension, but they lack explicit mechanisms to propagate low-level information from encoder to the decoder for improving segmentation performance. To overcome these challenges, we propose a Frequency-enhanced Vision Transformer (FEFormer) for robust and efficient volumetric medical image segmentation by explicitly model frequency information to jointly capture global context and fine structural details. Specifically, FEFormer employs a Frequency-enhanced Dynamic Self-Attention (FDSA) module that employs locality-preserving convolution with frequency-domain attention to jointly capture fine-grained local details and global long-range dependencies and dynamically integrates them based on frequency-domain importance. We further design a Frequency-decomposed Gating MLP (FGMLP) to adaptively disentangle and model low- and high-frequency components, enhancing both global semantic representation and local structural details. Additionally, a Wavelet-guided Adaptive Feature Fusion (WAFF) module is proposed to perform frequency-domain fusion of encoder and decoder features guided by Wavelet transformation, enabling semantically consistent and robust feature integration. Finally, we propose a Frequency-enabled Cross-scale Stem Bridge (FCSB) to enhance the propagation of low-level features from the encoder to the decoder by capturing their global representations and cross-scale interactions in the frequency domain. To demonstrate the effectiveness of FEFormer, we evaluated it on four diverse volumetric medical image segmentation tasks. Extensive experimental results demonstrated that FEFormer achieved superior segmentation performance with high computational efficiency compared to state-of-the-art methods.
\end{abstract}




\maketitle
\section{Introduction}
Segmentation of organs and lesions in medical images plays a critical role in clinical workflows, including diagnosis, prognosis, and treatment planning. However, manual delineation of target structures is labor-intensive, time-consuming, and prone to inter-observer variability, thereby motivating the development of automatic segmentation methods to improve efficiency and consistency \citep{wang2021annotation}. In recent years, deep learning (DL) techniques have significantly advanced this field, leading to the development of a wide range of automatic segmentation approaches \citep{azad2024medical}. Among these, convolutional neural networks (CNNs), particularly U-Net and its variants, have demonstrated remarkable performance by leveraging convolutional layers to effectively capture fine-grained local features \citep{ronneberger2015u,li2018h,yang2025dynamic,yang2025d2,yang2025dmc}. Nevertheless, due to the intrinsic locality of convolutional operations, CNN-based models are limited in their ability to capture global contextual information and model long-range dependencies \citep{yang2025translk,yang2026d}. Comparing with CNN-based models, the Vision Transformer (ViT) has been introduced as a powerful alternative backbone to capture global information for medical image segmentation \citep{dosovitskiy2021an}. It employs self-attention mechanisms to model long-range dependencies through global interactions across the entire image, enabling more effective utilization of contextual information. Due to these advantages, ViT-based methods have been widely adopted for organ and tissue segmentation across diverse imaging modalities, including Computed Tomography (CT) \citep{you2022class,yan2022after}, Cone Beam Computed Tomography \citep{chen2023cta}, Magnetic Resonance (MR) \citep{liu2022isegformer,pecco2024optimizing}, Positron Emission Tomography \citep{li2024swincross}, dermoscopy \citep{wu2022fat}, and electron microscopy \citep{pan2023adaptive}. However, their performance and applicability in medical image segmentation are challenged by the intrinsic characteristics of self-attention and multi-layer perceptrons (MLPs) modules and the architectural design of ViT-based networks.

\begin{figure*}[H]
\centering
\includegraphics[width=0.6\textwidth]{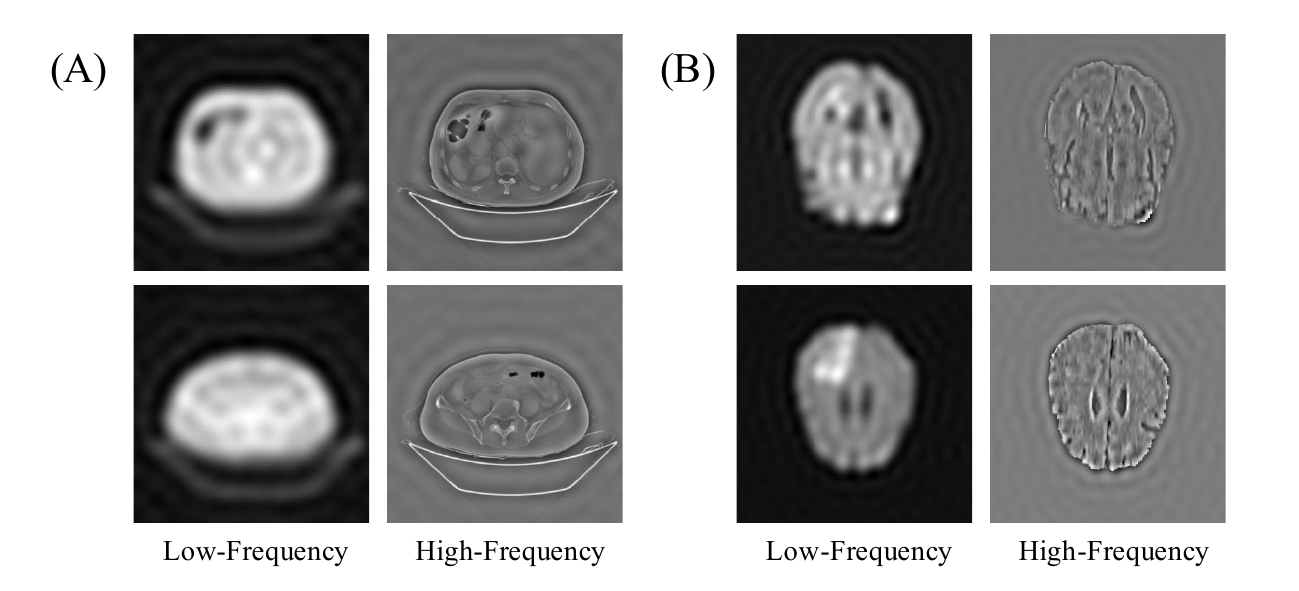}
\caption{Visualization of low-frequency and high-frequency components decomposed by 3D Fast Fourier Transformation on (A) CT volumes and (B) T1w MR volumes.}
\label{fig1}
\end{figure*}

First, ViT-based models leverage self-attention to model long-range spatial dependencies by computing pairwise interactions among tokenized image patches \citep{dosovitskiy2021an}. While this global modeling capability is effective for capturing contextual information, it may come at the cost of diminished sensitivity to fine-grained spatial details \citep{ren2022shunted,xu2025s2aformer}. Specifically, the global aggregation inherent in self-attention tends to lead to a bias toward low-frequency signals, thus ignoring high-frequency components \citep{zhang2023pha}. As shown in Fig.~\ref{fig1}, low-frequency components capture coarse structural information, while high-frequency components include information of peripheral regions and emphasize sharp intensity transitions and detailed patterns. Therefore, self-attention limits the model’s ability to preserve discriminative high-frequency information, thus diminishing representation of local structures such as edges, boundaries, and small anatomical variations \citep{bai2022improving,yan2025multi}. Moreover, standard self-attention operates primarily along the spatial dimension, focusing on relationships between patches while lacking explicit mechanisms to model inter-channel dependencies \citep{hatamizadeh2023global}. These self-attention mechanisms cannot fully capture cross-channel feature interactions to model channel attention due to the implementation of global pooling or aggregation operations to ignore the high-frequency information \citep{zheng2023lightweight}. Additionally, self-attention introduces substantial computational overhead due to the pairwise similarity computation between tokens \citep{liu2021swin,wu2025low}. For example, given an input with $N$ tokens, the attention operation incurs a computational complexity of $\mathcal{O}(N^3)$ which increases quadratically for high-resolution feature maps or volumetric medical images. This quadratic scaling not only increases memory consumption but also limits the practical deployment of ViT-based models in resource-constrained settings. 

Second, while ViT-based models employ MLPs to introduce non-linearity and enhance feature representation \citep{dosovitskiy2021an}, standard MLP modules operate in a point-wise manner across spatial locations and primarily model channel-wise transformations. Therefore, they lack explicit mechanisms to capture spatially localized patterns and fine-grained details. This limitation may lead to insufficient modeling of high-frequency spatial information, such as edges and small structures, which are essential for segmentation tasks. Additionally, conventional channel attention or dynamic weighting mechanisms typically rely on global pooling operations to summarize spatial information \citep{hu2018squeeze, wang2020eca, zhong2020squeeze}. While effective for capturing global context, such strategies predominantly encode low-frequency responses and may under-represent high-frequency components, thereby limiting their ability to fully exploit discriminative channel features.

Third, U-shaped segmentation architectures fuse skip-connected encoder features with upsampled decoder features through concatenation or element-wise summation to preserve spatial details and contextual information \citep{ronneberger2015u,cciccek20163d}. However, such fusion strategies treat features as homogeneous representations without aligning them and lack the ability to explicitly model their global relationships and semantic discrepancies \citep{song2026rmt}. To alleviate this limitation, several studies introduce attention mechanisms to selectively and adaptively fuse spatial features \citep{sun2025msm,xu2025x,yang2026d}. While effective, these approaches operate purely in the spatial domain. However, when features from encoder and decoder branches may exhibit substantial discrepancies in scale, semantics, and distribution, these approaches may lack semantic alignment and lead to unreliable fusion.

Fourth, most ViT-based segmentation models adopt a convolutional stem to generate overlapping patch embeddings by progressively downsampling the input images (e.g., to $\frac{H}{4}\times\frac{W}{4}$ resolution) \citep{xiao2021early,zhang2025lightweight}, and employ a symmetric stem in the decoder to restore spatial resolution (e.g., from $\frac{H}{4}\times\frac{W}{4}$ resolution) \citep{wang2022mixed,song2026rmt}. While this design enhances computational efficiency, it cannot fully capture low-level features and thus inevitably leads to the loss of fine-grained spatial details during early-stage downsampling. These low-level features, including edges, boundaries, and subtle texture variations, are particularly critical for accurate delineation in medical image segmentation. Since these ViT-based models lack mechanisms to propagate low-level features from the encoder stem to the decoder stem and fuse them accordingly \citep{cao2022swin}, existing ViT-based architectures often struggle to preserve anatomical precision and boundary fidelity, especially for small or low-contrast structures.

To address the aforementioned limitations of self-attention and MLP modules, we propose \textbf{Frequency-enhanced Dynamic Self-Attention} (FDSA) and \textbf{Frequency-decomposed Gating MLP} (FGMLP) modules. Specifically, FDSA first employs large-kernel depthwise convolution to capture rich fine-grained spatial details in large receptive fields, thus introducing a strong locality bias. This design enables the models to capture locally enriched features, thus improving the representation of edges, boundaries, and small anatomical structures for accurate medical image segmentation. Second, FDSA introduces a \textit{frequency-domain self-attention mechanism}, where self-attention is implemented on the frequency representations after spatial features are transformed into the spectral domain via the Fast Fourier Transform (FFT). Compared to spatial-domain attention, this design facilitates efficient modeling of large receptive fields, thus allowing the model to explicitly capture long-range dependencies through global frequency interactions. Additionally, global spatial attention often suppress high-frequency information related to fine-grained structures, but frequency-domain attention enables the preservation of discriminative high-frequency information. Third, FDSA incorporates a \textit{multi-frequency dynamic mechanism} to model inter-channel dependencies across distinct frequency bands. Instead of relying on a single global descriptor, the spectrum of features is decomposed into multiple frequency components, and channel-wise responses are adaptively modulated based on their spectral characteristics from different components separately. This design mitigates the bias toward low-frequency components and enables the network to preserve informative high-frequency signals, thereby capturing richer structural and textural patterns. These designs enable FDSA to maintain robustness across different imaging modalities and anatomical variations.

The FGMLP module first utilizes \textit{a gating mechanism} to preserve informative high-frequency signals which are associated with detailed structures such as boundaries and small regions. Furthermore, FGMLP incorporates a \textit{selective frequency decomposition mechanism} to adaptively modulates feature responses, improving segmentation performance across structures of varying scales. This mechanism decomposes input features into low-frequency and high-frequency components, and allows the model to disentangle global structural information from fine-grained details, facilitating more effective feature representation. It then selectively highlights informative frequency components, enhancing the network’s ability to capture both coarse semantic context and detailed spatial variations adaptively.

To address the limitations on feature fusion, we propose a \textbf{Wavelet-guided Adaptive Feature Fusion} (WAFF) module, which performs feature fusion in the frequency domain to enable semantically consistent and robust integration. Specifically, WAFF decomposes spatial features into multi-frequency components using the Discrete Wavelet Transform (DWT), where different subbands capture distinct semantic information, including global structures (low-frequency) and fine-grained details (high-frequency). By disentangling features into frequency-specific representations, WAFF performs adaptive fusion on corresponding subbands for aligning encoder and decoder features at the semantic level. This design ensures that components with similar semantic information are fused coherently, thereby mitigating the impact of large semantic discrepancies in the spatial domain. Furthermore, frequency-domain representations inherently capture global context with reduced spatial redundancy and improved noise suppression. After frequency decomposition, WAFF employs an adaptive fusion mechanism to enhance the robustness of feature fusion by emphasizing informative components while suppressing irrelevant or noisy signals. This mechanism models global relationships between encoder and decoder features within each frequency band, enabling dynamic and content-aware feature aggregation. Therefore, WAFF demonstrates more robust and efficient feature fusion and information aggregation without losing details and informative signals during fusion.

To address the limitations of low-level feature propagation, we propose a \textbf{Frequency-enabled Cross-scale Stem Bridge (FCSB)} to explicitly enhance the propagation and flow of low-level features between the stems of encoder and decoder. FCSB explores global dependencies and cross-scale interactions among varying-scaled features in the frequency domain. Specifically, it transforms spatial features into the frequency domain, and utilizes lightweight global receptive fields to capture long-range correlations among low-level features efficiently \citep{chi2020fast}. Additionally, it utilizes a frequency-guided cross-attention mechanism to explicitly model cross-scale interactions between shallow and deeper stem features in the frequency domain. By performing this cross-scale attention, FCSB effectively aligns multi-scale representations and enhances the consistency of structural details across resolutions. This design not only avoids fine-grained information lose during downsampling but also enables more effective integration of complementary features across scales. Therefore, FCSB enables low-level feature bridging by jointly exploiting frequency-domain global modeling and cross-scale attention with high computational efficiency.

Building upon those, we develop a \textbf{Frequency-enhanced Vision Transformer}, termed FEFormer, for generic knowledge extraction and adaptive feature fusion in robust and efficient volumetric medical image segmentation. We incorporate the FDSA and FGMLP into the Frequency-enhanced Transformer block by replacing standard self-attention and MLP modules, and incorporate the block into a hierarchical ViT architecture for adopting the scaling behavior of hierarchical transformers. We further integrate the WAFF into the decoder to fuse skip connected features and upsampled features, and integrate the FCSB between the convolutional stems of the encoder and the decoder. The incorporation of these modules jointly enhance spatial, spectral, and channel-wise feature representations, thereby enabling more accurate and robust medical image segmentation. We evaluated FEFormer on four heterogeneous volumetric segmentation tasks: abdominal multi-organ CT segmentation, brain tumor MR segmentation, hepatic vessel tumor CT segmentation, and abdomen organ CT segmentation. It achieved superior segmentation performance with lower computational complexity compared with other state-of-the-art (SOTA) models. Our contributions are summarized as follows:

\begin{itemize}
    \item We propose a \textbf{Frequency-enhanced Dynamic Self-Attention (FDSA)} module that integrates convolutions for strong locality bias with frequency-domain attention for efficient global context modeling. It further employs a multi-frequency dynamic mechanism to adaptively emphasize informative spectral components, improving the representation of fine-grained anatomical structures.
    \item We propose a \textbf{Frequency-decomposed Gating MLP (FGMLP)} module that combines gated modulation with frequency decomposition for adaptive enhancement of both global semantic context and fine structural details, leading to more discriminative feature representations. 
    \item We develop a \textbf{Wavelet-guided Adaptive Feature Fusion (WAFF)} module that performs frequency-domain feature decomposition and adaptive frequency-aware fusion to achieve semantically consistent integration and improve fusion robustness to large feature discrepancies.  
    \item We propose a \textbf{Frequency-enabled Cross-scale Stem Bridge (FCSB)} that leverages frequency-domain global modeling and cross-attention to enhance low-level feature propagation and improve consistency in structural fine details among multi-scale features.
    \item We develop a \textbf{Frequency-enhanced Vision Transformer (FEFormer)} by incorporating the proposed modules into a hierarchical ViT architecture for accurate, robust, and efficient medical image segmentation. We validated it on four diverse volumetric medical image segmentation tasks, demonstrating superior accuracy and high efficiency compared to SOTA methods.
\end{itemize}

\section{Related Works}
\subsection{Vision Transformers for Medical Image Segmentation}
ViT-based architectures have demonstrated remarkable success in medical image segmentation due to their strong capability in modeling long-range dependencies \citep{xiao2023transformers}. For example, Swin UNet represents one of the earliest fully transformer-based frameworks tailored for medical image segmentation, leveraging hierarchical self-attention to capture multi-scale contextual information \citep{cao2022swin}. Nevertheless, pure ViT-based models often struggle to effectively encode local structures due to their reliance on patch-wise tokenization. To address this limitation, Medical Transformer introduces gated position-sensitive axial attention along with a local–global training strategy to better balance fine-grained details and global context \citep{valanarasu2021medical}. Subsequent works have further explored hybrid and hierarchical designs to enhance both local and global feature modeling. MISSFormer incorporates enhanced transformer blocks and a context bridge module within a hierarchical architecture to facilitate cross-scale feature interaction \citep{huang2022missformer}. Similarly, the asymmetric compound branch Transformer adopts an asynchronous dual-branch design to efficiently capture complementary local and global dependencies while reducing computational complexity \citep{zhang2024ct}. To improve multi-scale representation learning, Dual Swin Transformer UNet employs dual-scale encoders to extract features at different semantic levels \citep{lin2022ds}, while nnFormer, built upon nnU-Net, integrates both local and global volume-based self-attention mechanisms to learn expressive volumetric representations \citep{zhou2023nnformer}. Despite these advancements, ViT-based models still exhibit limitations in capturing fine boundary and structural details, partly due to the smoothing effect of self-attention and MLP layers. To mitigate this issue, SEAformer enhances edge representation by explicitly preserving informative boundary features while suppressing background noise \citep{li2025seaformer}. Furthermore, the use of fixed-size patches restricts the ability of standard ViTs to model deformable anatomical structures. To overcome this constraint, AgileFormer introduces spatially dynamic components that adaptively adjust receptive fields, enabling more effective modeling of geometric variations and improving segmentation performance \citep{qiu2026agileformer}.

\subsection{Frequency in Vision Transformers}
The architecture of ViTs is enhanced by employing frequency filters and extracting frequency features. These methods highlight important frequency components for helping models focus on useful feature representations in target vision tasks. For instance, a Frequency Domain-based Transformer extracts features via frequency-based self-attention and MLP for efficient image deblurring \citep{kong2023efficient}. Similarly, the Holistic Dynamic Frequency Transformer is capable of capturing global information and dynamically selecting important frequency components for image fusion and restoration \citep{shang2024holistic}. Since high- and low-frequency components of features highlight different characteristics, the FreqFormer separately handles low- and high-frequency information for efficient image super-resolution \citep{dai2024freqformer}. The LoFormer simultaneously captures features from low- and high-frequency local windows, thus effectively modeling long-range dependencies with maintaining fine-grained details for image deblurring \citep{mao2024loformer}. Other methods employ frequency mechanisms to extract more features in the frequency domain, thus enhancing and supplement the extraction of spatial features to improve performance in vision tasks. For example, the SpectFormer employs frequency attention layers to enhance spatial self-attention layers by capturing high-frequency information to improve segmentation in edges and lines \citep{patro2025spectformer}. The frequency feature aggregation transformer captures spatial features to enhance high-frequency feature extraction in image super-resolution effectively \citep{song2025efficient}. The DBFFT employs a dual frequency and spatial branches to learn representation from the frequency and spatial domains simultaneously and to adaptively fuse features from two domains \citep{zeng2024dbfft}.

\subsection{Wavelet Transform in Semantic Segmentation}
Wavelet transform offers a unique advantage in jointly modeling spatial and frequency information, making it particularly well-suited for semantic segmentation tasks where both global structures and fine-grained details are critical. Consequently, it has been increasingly incorporated into segmentation frameworks. For instance, WNet leverages a wavelet-based encoder to capture cross-modal representations from video and audio signals, enabling more effective object segmentation \citep{pan2022wnet}. XNet decomposes biomedical images into low- and high-frequency components via wavelet transform, facilitating consistency learning across frequency bands and thereby improving robustness \citep{zhou2023xnet}. Similarly, WCMamba introduces a pyramid wavelet mechanism to extract multi-scale representations and enhance fine structural details, leading to improved segmentation performance \citep{zhan2025wcmamba}. WTCLIP further integrates a learnable wavelet transform decoder to strengthen feature extraction capabilities within CLIP, particularly benefiting boundary delineation \citep{xiao2026wtclip}. In the context of remote sensing, FSSFFormer employs discrete wavelet transform to decompose images into frequency sub-bands, explicitly preserving global spatial structures while maintaining high-frequency details, which is essential for accurate semantic segmentation \citep{li2026frequency}.

\section{Methods}
\subsection{Overall Architecture}
\begin{figure*}[!t]
\centering
\includegraphics[width=\textwidth]{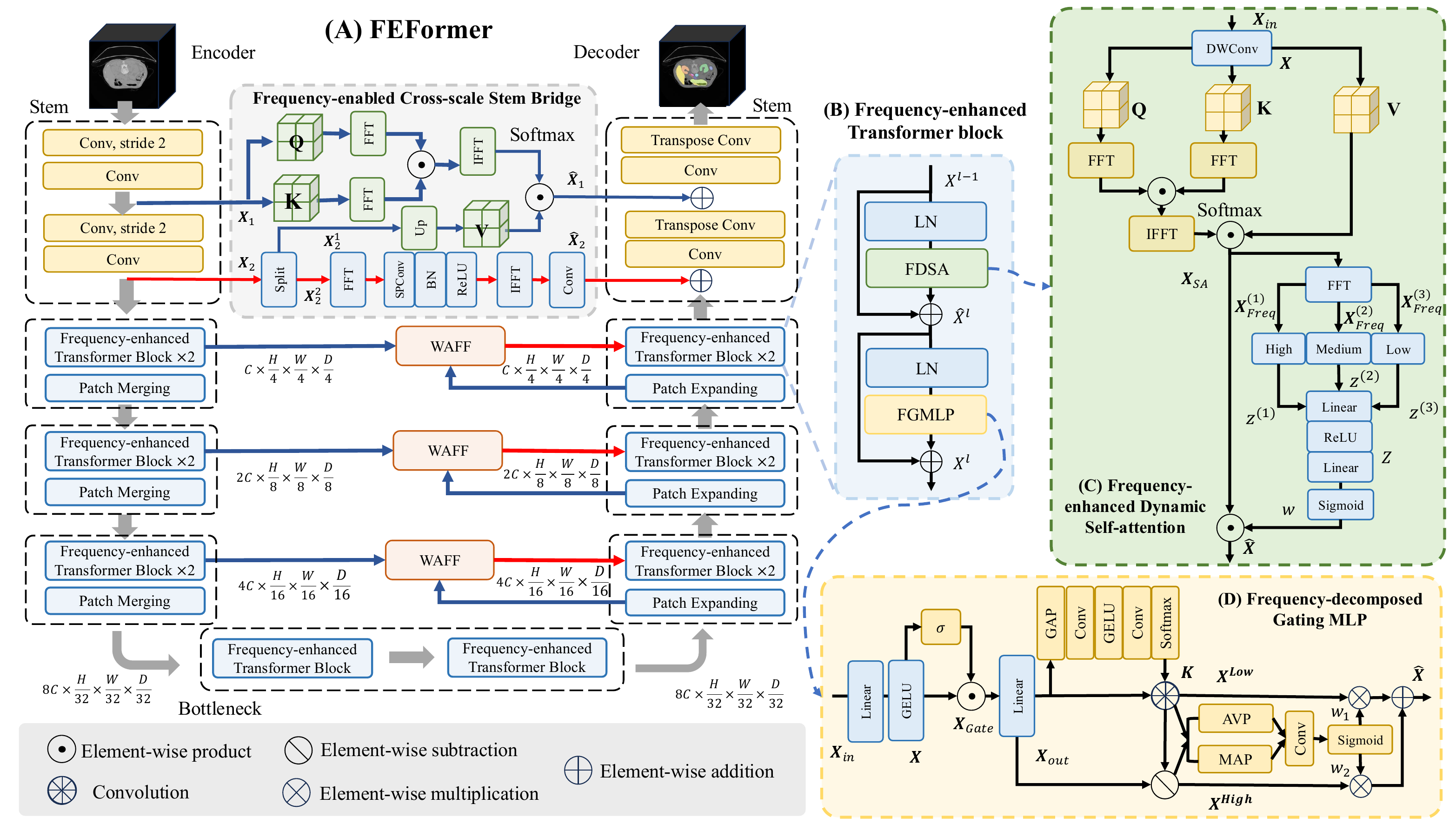}
\caption{(A) The overall architecture of FEFormer. FEFormer consists of an encoder, a bottleneck, a decoder, and a Frequency-enabled Cross-scale Stem Bridge (FCSB). The encoder employs a convolutional stem, and subsequently employs two consecutive Frequency-enhanced Transformer blocks and a patch merging layer at each stage. The decoder with a symmetry architecture employs a patch expanding layer and two Transformer blocks at each stage. The stem bridge employs FCSB to propagate features from the encoder to the decoder. (B) The Frequency-enhanced Transformer block incorporates a FDSA and FGMLP modules. (C) The Frequency-enhanced Dynamic Self-attention module utilizes a frequency-domain self-attention mechanisms and a multi-frequency dynamic mechanism. (D) The Frequency-decomposed gating MLP module utilizes a gating mechanism and a selective frequency decomposition mechanism.}
\label{fig2}
\end{figure*}

FEFormer adopts a U-shaped encoder–decoder architecture to learn hierarchical feature representations (Fig.~\ref{fig2}). It consists of four main components: an encoder, a bottleneck, a decoder, and a stem bridge.

\textbf{Encoder.} The encoder employs a convolutional stem that downsamples the input volume of size $C_{in} \times D \times H \times W$ by a factor of 4, projecting it into a $C$-dimensional feature space ($C=64$) and generating patches with size $C \times \frac{H}{4} \times \frac{W}{4} \times \frac{D}{4}$. It comprises three stages, each containing two consecutive Frequency-Enhanced Transformer blocks. After transformer blocks at each stage, a patch merging layer employs a $3 \times 3 \times 3$ convolution with stride 2 followed by layer normalization to progressively reduce the spatial resolution while doubling the channel dimension. Therefore, the feature maps at each stage are of sizes $C \times \frac{H}{4} \times \frac{W}{4}\times \frac{D}{4}$, $2C \times \frac{H}{8} \times \frac{W}{8} \times \frac{D}{8}$, and $4C \times \frac{H}{16} \times \frac{W}{16} \times \frac{D}{16}$, respectively.

\textbf{Bottleneck.} The bottleneck further processes the deepest features using two consecutive Frequency-Enhanced Transformer blocks, producing representations of size $8C \times \frac{H}{32} \times \frac{W}{32} \times \frac{D}{32}$.

\textbf{Decoder.} The decoder utilizes a similar architecture as the encoder with three stages, each consisting of two consecutive Frequency-Enhanced Transformer blocks. Before each stage, a patch expanding layer is employed to upsample the feature maps by a factor of 2. The upsampled features are then fused with the corresponding skip-connected encoder features adaptively within a WAFF block. Consequently, the feature dimensions at the decoder stages are $4C \times \frac{H}{16} \times \frac{W}{16} \times \frac{D}{16}$, $2C \times \frac{H}{8} \times \frac{W}{8} \times \frac{D}{8}$, and $C \times \frac{H}{4} \times \frac{W}{4} \times \frac{D}{4}$, respectively. Subsequently, a convolution stem is employed to reconstruct the feature maps by restoring the original spatial resolution of input images. Lastly, a $1\times1\times1$ convolutional layer is employed to produce the voxel-wise segmentation predictions.

\textbf{Stem bridge.} A FCSB is employed to bridge the feature flows between the convolution stems of the encoder and decoder, enhancing low-level feature propagation and improving the preservation of fine structural details.

\subsection{Frequency-enhanced Transformer block}
The Frequency-enhanced Transformer block is built on the FDSA and FGMLP modules (Fig.~\ref{fig2}). Layer Normalization (LN) layer is applied before the FDSA and FGMLP modules, with residual connections incorporated in both. Given the input in the $l-$th layer as $\boldsymbol{X}^{l-1}\in\mathbb{R}^{C\times H \times W\times D}$, the output, $\boldsymbol{X}^l\in\mathbb{R}^{C\times H \times W\times D}$, is computed as
\begin{align}
    \hat{\boldsymbol{X}}^l&=\textrm{FDSA}(\textrm{LN}(\boldsymbol{X}^{l-1}))+\boldsymbol{X}^{l-1}, \\
    \boldsymbol{X}^l&=\textrm{FGMLP}(\textrm{LN}(\boldsymbol{\hat{\boldsymbol{X}}^l}))+\hat{\boldsymbol{X}}^l.
\end{align}

\subsection{Frequency-enhanced Dynamic Self-attention module}
The FDSA module is constructed by integrating a \textit{frequency-domain self-attention mechanism} into a \textit{multi-frequency dynamic mechanism} (Fig.~\ref{fig2}). Specifically, the \textit{frequency-domain self-attention mechanism} employs a depth-wise convolutional layer with a large $7\times7\times7$ kernel ($\textrm{DWConv}_{7\times7\times7}$) to extract features $\boldsymbol{X}\in\mathbb{R}^{C\times H\times W\times D}$ from the input $\boldsymbol{X}_{\textrm{in}}\in\mathbb{R}^{C\times H\times W\times D}$ as
\begin{align}
    \boldsymbol{X}=\textrm{DWConv}_{7\times7\times7}(\boldsymbol{X}_{\textrm{in}}).
\end{align}
This large-kernel depth-wise convolutional layer is employed to capture rich spatial features through an enlarged receptive field while maintaining low computational complexity. Additionally, it introduces local inductive biases into the self-attention mechanism, thereby enhancing the model’s ability to efficiently capture fine-grained features and spatial details. These spatial features are projected for generating the query $\boldsymbol{Q}$, the key $\boldsymbol{K}$, and the value $\boldsymbol{V}$ via the linear matrices as
\begin{align}
    \boldsymbol{Q}=\boldsymbol{X}\boldsymbol{W}_q,\; \boldsymbol{K}=\boldsymbol{X}\boldsymbol{W}_k,\; \boldsymbol{V}=\boldsymbol{X}\boldsymbol{W}_v.
\end{align}
The query and key are converted to frequency-domain features via FFT as
\begin{align}
    \boldsymbol{Q}_{\textrm{Freq}}=\textrm{FFT}(\boldsymbol{Q}),\;\boldsymbol{K}_{\textrm{Freq}}=\textrm{FFT}(\boldsymbol{K}).
\end{align}
Subsequently, frequency-domain attention score $\boldsymbol{QK}_{\textrm{Freq}}$ is calculated as
\begin{align}
    \boldsymbol{QK}_{\textrm{Freq}}=\boldsymbol{Q}_{\textrm{Freq}}\odot\boldsymbol{K}_{\textrm{Freq}}.
\end{align}
Subsequently, the frequency-domain attention score is converted to the spatial domain by implementing the Inverse Fast Fourier Transform (IFFT) as
\begin{align}
    \boldsymbol{QK}=\textrm{IFFT}(\boldsymbol{QK}_{\textrm{Freq}}).
\end{align}
The output features of the \textit{frequency-domain self-attention mechanism} $\boldsymbol{X}_{\textrm{SA}}\in\mathbb{R}^{C\times H\times W\times D}$ is generated by multiplying the attention score and the value as
\begin{align}
    \boldsymbol{X}_{\textrm{SA}}=\textrm{Softmax}(\boldsymbol{QK})\cdot\boldsymbol{V}.
\end{align}
Standard spatial-domain self-attention may ignore high-frequency information because its global aggregation operation tends to smooth the fine-grained features and local patterns captured by large-kernel convolutions, thus limiting segmentation performance in edges and boundary structures. In contrast, this \textit{frequency-domain self-attention mechanism} calculates global self-attention scores through element-wise multiplication in the frequency domain. According to the Convolution Theorem, element-wise multiplication in the frequency domain is equivalent to convolution in the spatial domain. Therefore, the proposed mechanism explicitly preserves high-frequency information rather than allowing it to be implicitly smoothed or diluted during spatial-domain aggregation. This enables the model to better retain fine-grained features that are critical for accurate segmentation of edges, boundaries, fine anatomical structures, and small abnormalities. Furthermore, this \textit{frequency-domain self-attention mechanism} reduces the computational complexity to $O(N\log N)$, significantly reducing from $O(N^2)$in the standard self-attention.

Subsequently, a \textit{multi-frequency dynamic mechanism} is introduced to adaptively recalibrate channel features according to their importance in the frequency domain. By explicitly analyzing the contributions of low-, medium-, and high-frequency components, this mechanism assigns larger weights to channels containing discriminative global structures, texture patterns, or fine-grained details, while reducing the influence of redundant or noisy responses. This dynamic recalibration enables the network to capture complementary information across multiple frequency bands and improves feature representation quality. 

Existing channel attention mechanisms typically employ global average pooling to estimate channel importance. However, average pooling primarily reflects low-frequency responses and may overlook informative mid- and high-frequency signals that are critical for representing boundaries, textures, and small anatomical structures. In contrast, this \textit{multi-frequency dynamic mechanism} decomposes feature representations into multiple frequency sub-bands and models their contributions separately. This frequency-aware representation enables a more comprehensive characterization of channel-wise information and allows the network to adaptively emphasize complementary features across different frequency ranges. 

Specifically, the features $\boldsymbol{X}_{\textrm{SA}}$ along all channels $C$ are transformed to the frequency domain via FFT as
\begin{align}
    \boldsymbol{X}_{\textrm{Freq}}=\textrm{FFT}(\boldsymbol{X}_{\textrm{SA}}).
\end{align}
This frequency-domain features $\boldsymbol{X}_{\textrm{Freq}}$ are decomposed into three sub-band components $\boldsymbol{X}^{(s)}_{\textrm{Freq}}$ ($s\in\{1,2,3\}$), including low-frequency $\boldsymbol{X}^{(1)}_{\textrm{Freq}}$, mid-frequency $\boldsymbol{X}^{(2)}_{\textrm{Freq}}$, and high-frequency $\boldsymbol{X}^{(3)}_{\textrm{Freq}}$, as
\begin{align}
    \{\boldsymbol{X}^{(1)}_{\textrm{Freq}},\boldsymbol{X}^{(2)}_{\textrm{Freq}},\boldsymbol{X}^{(3)}_{\textrm{Freq}}\}=\{\boldsymbol{X}^{(s)}_{\textrm{Freq}}\}^3_{s=1}=\textrm{Decomp}(\boldsymbol{X}_{\textrm{Freq}}).
\end{align}
These spectrum components are globally aggregated within the spatial region $\Omega$ of each sub-band to model their relative importance as
\begin{align}
    \{z^{(1)},z^{(2)},z^{(3)}\}=\{z^{(s)}\}^3_{s=1}=\frac{1}{|\Omega|}\sum_{(d,h,w)\in\Omega}|\boldsymbol{X}^{(s)}_{\textrm{Freq}}|.
\end{align}
All aggregated frequency components are concatenated along channels to generate frequency-domain descriptors $\boldsymbol{Z}$ to model channel-wise inter-dependencies as
\begin{align}
    \boldsymbol{Z}=\textrm{Concate}([z^{(1)};z^{(2)};z^{(3)}]).
\end{align}
Subsequently, the dynamic mechanism is employed to generate channel-wise dynamic weight $w$ by stacking two fully connected (FC) layers, with a ReLU function in-between and a sigmoid function followed as
\begin{align}
    w=\textrm{Sigmoid}\big(\textrm{FC}\big(\textrm{ReLU}(\textrm{FC}(\boldsymbol{Z}))\big)\big).
\end{align}
The channel-wise inter-dependencies are modeled based on frequency signals. The dynamically calibrated features $\hat{\boldsymbol{X}}\in\mathbb{R}^{C\times H\times W\times D}$ are generated by adaptively highlighting the important and informative features by the channel-wise dynamic weight as
\begin{align}
    \hat{\boldsymbol{X}}=w\odot\boldsymbol{X}_{\textrm{SA}}.
\end{align}

\subsection{Frequency-decomposed Gating MLP module}
The Frequency-decomposed Gating MLP (FGMLP) module is built by integrating a \textit{gating mechanism} with a \textit{selective frequency decomposition mechanism} (Fig.~\ref{fig2}). Specifically, given the input features $\boldsymbol{X}_{in}\in\mathbb{R}^{C\times H\times W\times D}$, a linear layer is employed to project them to features $\boldsymbol{X}\in\mathbb{R}^{4C\times H\times W\times D}$ by expanding channels with a ratio of 4 with a GELU followed as
\begin{align}
    \boldsymbol{X}&=\textrm{GELU}(\textrm{Linear}(\boldsymbol{X}_{in})).
\end{align}
Subsequently, the \textit{gating mechanism} generates gated features $\boldsymbol{X}_{\textrm{Gate}}\in\mathbb{R}^{4C\times H\times W\times D}$ by utilizing an activation function $\sigma$  (e.g., $\textrm{ReLU6}$) as
\begin{align}
    \boldsymbol{X}_{\textrm{Gate}}=\boldsymbol{X}\odot\sigma(\boldsymbol{X}).
\end{align}
Channel-wise MLP module lacks mechanisms to extract fine grained spatial features. However, employing a gating mechanism enables it to enhance the extraction of spatial features and highlight fine grained details by amplifying high frequency signals.

After gating, a second linear projection layer is employed to compress channels to the original dimension as $\boldsymbol{X}_{out}\in\mathbb{R}^{C\times H\times W\times D}$, and two Dropout layers are employed before and after it as
\begin{align}
    \boldsymbol{X}_{out}=\textrm{Dropout}(\textrm{Linear}(\textrm{Dropout}(\boldsymbol{X}_{\textrm{Gate}}))).
\end{align}
Subsequently, the \textit{selective frequency decomposition mechanism} decomposes the features $\boldsymbol{X}_{out}$ into low-frequency and high-frequency components using learnable low frequency filters and adaptively highlights important ones. Specifically, input-adaptive low-pass frequency filter kernels are first generated by stacking a global average pooling layer ($\textrm{GAP}$), two $1\times1\times1$ convolutional layer ($\textrm{Conv}$) with a GELU function in-between, and a softmax function. The global average pooling layer is utilized to describe global contextual information, and the $1\times1\times1$ convolutional layers project global context with a reduction ratio of 4 in intermediate channel dimensions for generating filter kernels with the size of $k^3$. To enhance stability, the softmax function is utilized to normalize kernels as low-pass smoothing filters. Thus, the kernels $\boldsymbol{K}\in\mathbb{R}^{C\times k^3}$ are generated as
\begin{align}
    \boldsymbol{K}=\textrm{Softmax}\big(\textrm{Conv}\big(\textrm{GELU}\big(\textrm{Conv}\big(\textrm{GAP}(\boldsymbol{X}_{\textrm{out}})\big)\big)\big)\big).
\end{align}
The normalized kernels are reshaped into convolutional filters $\boldsymbol{K} \in \mathbb{R}^{C \times 1 \times 1 \times 1 \times k^3}$, and this input-adaptive kernel is applied to extract low-frequency components via a convolutional layer as
\begin{align}
    \boldsymbol{X}^{\textrm{Low}}=\textrm{Conv}(\boldsymbol{K},\boldsymbol{X}_{\textrm{out}}).
\end{align}
The high-frequency components $\boldsymbol{X}^{\textrm{High}}$ are extracted from the input features by removing low-frequency components from the whole signals as
\begin{align}
    \boldsymbol{X}^{\textrm{High}}=\boldsymbol{X}_{\textrm{out}}-\boldsymbol{X}^{\textrm{Low}}.
\end{align}
Aggregating information within spatial regions preserves low-frequency signals while suppressing high-frequency components, as spatial averaging smooths fine-grained details in the feature maps. Therefore, the low-pass frequency filter is constructed using a global average pooling layer followed by $1\times1\times1$ convolutional layers.

A channel-wise frequency selection modulator is then introduced to adaptively emphasize informative frequency components from the low- and high-frequency branches, thereby enhancing discriminative channel features according to their relevance. Since different channels may rely on distinct frequency characteristics, dynamically reweighting these components allows the network to better capture global structural information from low frequencies and fine boundary details from high frequencies. To estimate channel importance robustly, both average pooling ($\textrm{AVP}$) and max pooling ($\textrm{MAP}$) layers are employed to summarize complementary global responses, where average pooling captures overall activation statistics and max pooling highlight salient responses. These descriptors are then interacted through a $7\times7\times7$ depth-wise convolutional layer ($\textrm{DWConv}_{7\times7\times7}$) kernel and sigmoid activation to generate adaptive channel-wise modulation weights as
\begin{align}
    w_{avp}&=\textrm{AVP}(\boldsymbol{X}^{\textrm{Low}}+\boldsymbol{X}^{\textrm{High}}),\\
    w_{map}&=\textrm{MAP}(\boldsymbol{X}^{\textrm{Low}}+\boldsymbol{X}^{\textrm{High}}),\\
    \{w_1,w_2\}&=\textrm{Sigmoid}\big(\textrm{DWConv}_{7\times7\times7}([w_{avp};w_{map}])\big).
\end{align}
Lastly, the output of the FGMLP module $\hat{\boldsymbol{X}}\in\mathbb{R}^{C\times H\times W\times D}$ is generated by adaptively re-weighting low-frequency and high-frequency features based on their relative importance as
\begin{align}
    \hat{\boldsymbol{X}} = w_1 \odot \boldsymbol{X}^{\textrm{Low}} + w_2 \odot\boldsymbol{X}^{\textrm{High}}.
\end{align}

\subsection{Wavelet-guided Adaptive Feature Fusion module}
\begin{figure*}[!t]
\centering
\includegraphics[width=0.7\textwidth]{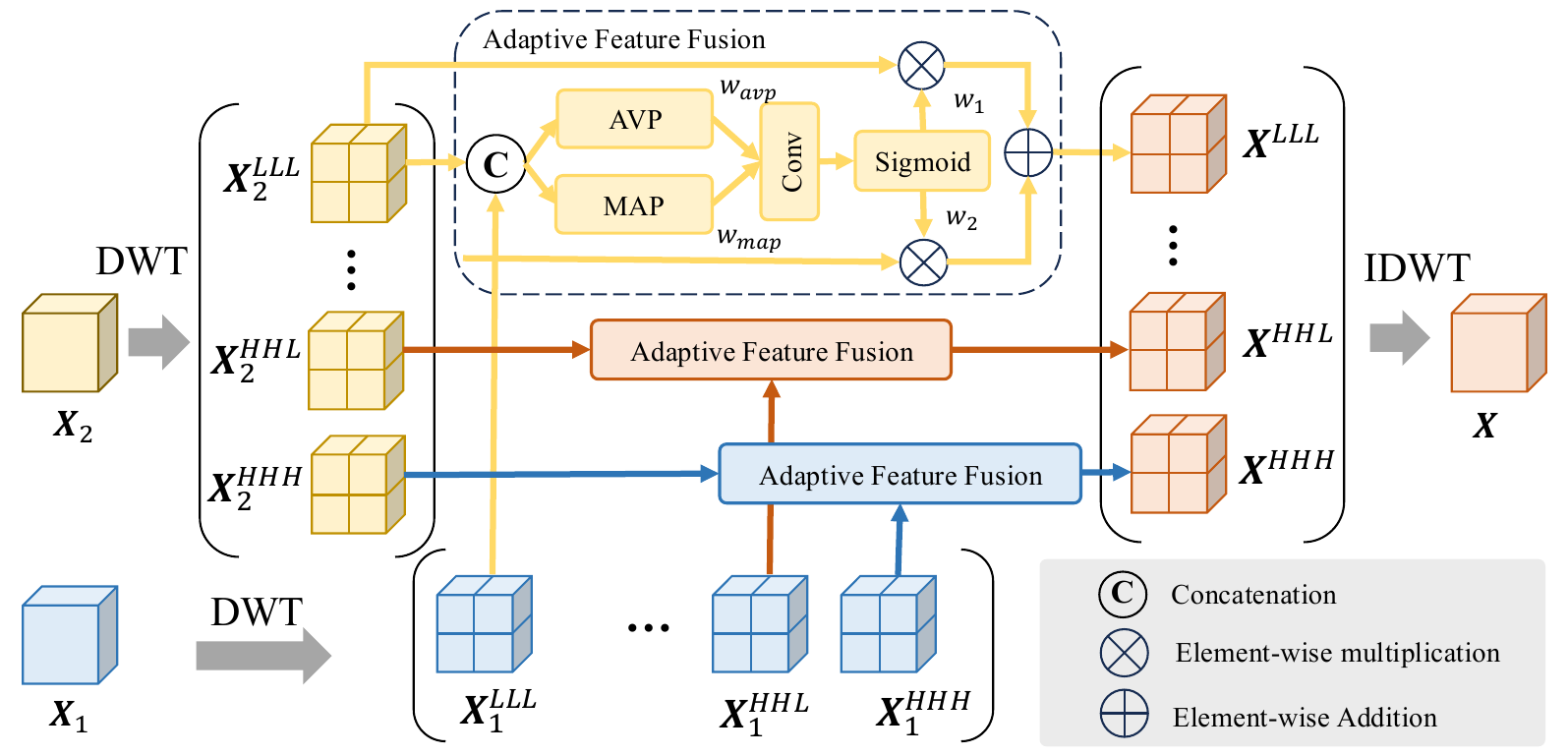}
\caption{The architecture of the Wavelet-guided Adaptive Feature Fusion (WAFF) module. WAFF takes input spatial features $\boldsymbol{X}_1$ and $\boldsymbol{X}_2$, and employs DWT to decompose them into frequency-domain sub-bands: $\{\boldsymbol{X}_1^{LLL},...,\boldsymbol{X}_1^{HHL},\boldsymbol{X}_1^{HHH}\}$ and $\{\boldsymbol{X}_2^{LLL},...,\boldsymbol{X}_2^{HHL},\boldsymbol{X}_2^{HHH}\}$. Subsequently, an adaptive feature fusion mechanism is employed to fuse each corresponding sub-bands. All fused sub-bands $\{\boldsymbol{X}^{LLL},...,\boldsymbol{X}^{HHL},\boldsymbol{X}^{HHH}\}$ are converted back to the spatial domain via IDWT as output features $\boldsymbol{X}$.} 
\label{fig3}
\end{figure*}

We propose a Wavelet-guided Adaptive Feature Fusion (WAFF) module for semantically consistent feature fusion in the frequency domain, thus improving fusion robustness by aligning feature semantics to handle large feature discrepancy (Fig.~\ref{fig3}). Specifically, WAFF employs Haar Discrete Wavelet Transform (DWT) to decompose the input features $\boldsymbol{X}_1$ and $\boldsymbol{X}_2$ ($\boldsymbol{X}_1,\boldsymbol{X}_2\in\mathbb{R}^{C\times H\times W\times D}$) into eight sub-bands in the frequency domain by separating low-frequency and high-frequency signals along three dimensions $\{\boldsymbol{X}_1^{\textrm{LLL}},\boldsymbol{X}_1^{\textrm{LLH}},...,\boldsymbol{X}_1^{\textrm{HHH}}\}\in\mathbb{R}^{C\times\frac{H}{2}\times\frac{W}{2}\times\frac{D}{2}}$ and $\{\boldsymbol{X}_2^{\textrm{LLL}},\boldsymbol{X}_2^{\textrm{LLH}},...,\boldsymbol{X}_2^{\textrm{HHH}}\}\in\mathbb{R}^{C\times\frac{H}{2}\times\frac{W}{2}\times\frac{D}{2}}$, respectively, as
\begin{align}
    \{\boldsymbol{X}_1^{\textrm{LLL}},\boldsymbol{X}_1^{\textrm{LLH}},...,\boldsymbol{X}_1^{\textrm{HHH}}\}&=\textrm{DWT}(\boldsymbol{X}_1), \\
    \{\boldsymbol{X}_2^{\textrm{LLL}},\boldsymbol{X}_2^{\textrm{LLH}},...,\boldsymbol{X}_2^{\textrm{HHH}}\}&=\textrm{DWT}(\boldsymbol{X}_2).
\end{align}
Each sub-band represents different semantics, and take $\boldsymbol{X}_1$ for example:
\begin{itemize}
    \item Sub-band $\boldsymbol{X}_1^{\textrm{LLL}}$ represents pure low-frequency semantic features.
    \item Sub-bands $\{\boldsymbol{X}_1^{\textrm{LLH}},\boldsymbol{X}_1^{\textrm{LHL}},\boldsymbol{X}_1^{\textrm{HLL}}\}$ represent relatively higher-frequency semantic features related to edges and boundaries along three dimensions $(D,W,H)$. 
    \item Sub-bands $\{\boldsymbol{X}_1^{\textrm{LHH}},\boldsymbol{X}_1^{\textrm{HLH}},\boldsymbol{X}_1^{\textrm{HHL}}\}$ represent higher-frequency semantic features related to more complex structures.
    \item Sub-band $\boldsymbol{X}_1^{\textrm{HHH}}$ represents high-frequency semantic features related to very fine details or noise.
\end{itemize}
Subsequently, an adaptive feature fusion mechanism is utilized to fuse corresponding sub-bands, thus aligning their semantics in the frequency domain. We take $\boldsymbol{X}_1^{\textrm{LLL}}$ and $\boldsymbol{X}_2^{\textrm{LLL}}$ for example. Their global relationship description is efficiently modeled by applying average pooling ($\textrm{AVP}$) and max pooling ($\textrm{MAP}$) along channels as
\begin{align}
    w_{avp} &= \textrm{AVP}(\textrm{Concate}([\boldsymbol{X}_1^{\textrm{LLL}};\boldsymbol{X}_2^{\textrm{LLL}}])), \\
    w_{map} &= \textrm{MAP}(\textrm{Concate}([\boldsymbol{X}_1^{\textrm{LLL}};\boldsymbol{X}_2^{\textrm{LLL}}])).
\end{align}
Subsequently, a $7\times7\times7$ depthwise convolutional layer ($\textrm{DWConv}_{7\times7\times7}$) is used to allow such information to interact and mix among two semantic descriptors. Lastly, a Sigmoid function is used to obtain dynamic selection values $w_1$, $w_2$ as
\begin{align}
\nonumber
    \{w_1;w_2\} &= \textrm{Sigmoid}(\textrm{DWConv}_{7\times7\times7}([w_{avp};w_{map}])).
\end{align}
Two sub-bands are calibrated by these selection weights and adaptively fused as
\begin{align}
    \boldsymbol{X}^{\textrm{LLL}} &=w_1 \odot \boldsymbol{X}_1^{\textrm{LLL}} + w_2 \odot \boldsymbol{X}_2^{\textrm{LLL}} .
\end{align}
WAFF decomposes spatial features into multiple frequency-domain sub-bands to capture complementary semantic information, where low-frequency components encode global structural context while high-frequency components preserve fine-grained details. This frequency-aware decomposition enables spatial patterns to be represented according to their underlying frequency characteristics, thereby facilitating semantic disentanglement across different feature components. Consequently, WAFF performs alignment and fusion on sub-features with similar semantic properties, avoiding the mismatch introduced by substantial spatial discrepancies. Furthermore, an adaptive feature fusion mechanism dynamically emphasizes informative components while suppressing irrelevant or noisy responses, enabling more effective feature aggregation without compromising structural details.

All fused sub-bands are combined and transformed back to the spatial domain via Inverse Discrete Wavelet Transform (IDWT), thus generating the output $\boldsymbol{X}\in\mathbb{R}^{C\times H\times W\times D}$ as
\begin{align}
    \boldsymbol{X}=\textrm{IDWT}(\boldsymbol{X}^{\textrm{LLL}}, \boldsymbol{X}^{\textrm{LLH}},..., \boldsymbol{X}^{\textrm{HHL}}, \boldsymbol{X}^{\textrm{HHH}}).
\end{align}

\subsubsection{Frequency-enabled Cross-scale Stem Bridge}
The stem in the encoder consists of two patch embedding layers, and each patch embedding layer employs two successive $3\times3\times3$ convolutional layers with a stride of 2 and 1 ($\textrm{Conv}_{3\times3\times3,2}$ and $\textrm{Conv}_{3\times3\times3,1}$, respectively). A Batch Normalization and a GELU activation function are employed after each convolutional layer. Two patch embedding layers project the channel number of the input image $\boldsymbol{X}_{in}\in\mathbb{R}^{C_{in}\times H\times W\times D}$ to $C/2$ and $C$ channels, respectively, thus generating the output features of two patch embedding layers $\boldsymbol{X}_1\in\mathbb{R}^{\frac{C}{2}\times\frac{H}{2}\times\frac{W}{2}\times\frac{D}{2}}$ and $\boldsymbol{X}_2\in\mathbb{R}^{C\times\frac{H}{4}\times\frac{W}{4}\times\frac{D}{4}}$ as
\begin{align}
    \boldsymbol{X}_1&=\textrm{GELU}(\textrm{BN}(\textrm{Conv}_{3\times3\times3,2}(\boldsymbol{X}_{in}))), \\
    \boldsymbol{X}_1&=\textrm{GELU}(\textrm{BN}(\textrm{Conv}_{3\times3\times3,1}(\boldsymbol{X}_1))), \\
    \boldsymbol{X}_2&=\textrm{GELU}(\textrm{BN}(\textrm{Conv}_{3\times3\times3,2}(\boldsymbol{X}_1))), \\
    \boldsymbol{X}_2&=\textrm{GELU}(\textrm{BN}(\textrm{Conv}_{3\times3\times3,1}(\boldsymbol{X}_2))).
\end{align}
To enhance the low-level feature propagation of the stem from the encoder to the decoder, we propose a Frequency-enabled Cross-scale Stem Bridge (FCSB) (Fig.~\ref{fig2}). The output features $\boldsymbol{X}_2$ is split into two subset of features along channels as $\boldsymbol{X}_2^1$ and $\boldsymbol{X}_2^2$ ($\boldsymbol{X}_2^1,\boldsymbol{X}_2^2\in\mathbb{R}^{\frac{C}{2}\times\frac{H}{4}\times\frac{W}{4}\times\frac{D}{4}}$). Subsequently, the efficient global feature interaction is performed to the sub-features $\boldsymbol{X}_2^2$ in the frequency domain for efficiently global context modeling. Specifically, the spatial features $\boldsymbol{X}_2^2$ are transformed to frequency representations via FFT. A spectral $1\times1\times1$ convolutional layer ($\textrm{SPConv}_{1\times1\times1}$), Batch Normalization (BN), and ReLU are then stacked and employed as a global receptive field on real and imaginary parts to learn global representations. These frequency representations are transformed back to the spatial domain for capturing globally enriched feature embeddings while preserving structural details. A $1\times1\times1$ convolutional layer is then employed to restore the channel number of spatial features $\hat{\boldsymbol{X}}_2\in\mathbb{R}^{C\times\frac{H}{4}\times\frac{W}{4}\times\frac{D}{4}}$ as
\begin{align}
    \boldsymbol{X}_{2,\textrm{Freq}}&=\textrm{FFT}(\boldsymbol{X}_2^2), \\
    \hat{\boldsymbol{X}}_{2,\textrm{Freq}}&=\textrm{ReLU}\big(\textrm{BN}(\textrm{SPConv}_{1\times1\times1}(\boldsymbol{X}_{2,\textrm{Freq}}))\big), \\
    \hat{\boldsymbol{X}}_2&=\textrm{Conv}(\textrm{iFFT}(\hat{\boldsymbol{X}}_{2,\textrm{Freq}})).
\end{align}
Another subset of features $\boldsymbol{X}_2^1$ are upsampled to the same dimension as that of features from the first patch embedding layer $\boldsymbol{X}_1$. Subsequently, their cross-scale correlations are captured by the frequency-domain cross-attention mechanism for generating features $\hat{\boldsymbol{X}}_1\in\mathbb{R}^{\frac{C}{2}\times\frac{H}{2}\times\frac{W}{2}\times\frac{D}{2}}$ as
\begin{align}
    \boldsymbol{X}_2^1&=\textrm{Upsample}(\boldsymbol{X}_2^1), \\
    \boldsymbol{Q}&=\boldsymbol{X}_1\boldsymbol{W}_q,\; \boldsymbol{K}=\boldsymbol{X}_1\boldsymbol{W}_k,\;\boldsymbol{V}=\boldsymbol{X}_2^1\boldsymbol{W}_v\\
    \boldsymbol{Q}_{\textrm{Freq}}&=\textrm{FFT}(\boldsymbol{Q}),\; \boldsymbol{K}_{\textrm{Freq}}=\textrm{FFT}(\boldsymbol{K}),\\
    \boldsymbol{QK}&=\textrm{iFFT}(\boldsymbol{Q}_{\textrm{Freq}}\odot\boldsymbol{K}_{\textrm{Freq}}),\\
    \hat{\boldsymbol{X}}_1&=\textrm{Softmax}(\boldsymbol{QK})\cdot\boldsymbol{V}.
\end{align}
Therefore, this frequency-domain cross-attention mechanism captures global cross-scale dependencies between shallow high-resolution features and deep semantic representations. This feature correlation modeling in the spectral domain enables efficient long-range interaction and semantic-guided feature refinement for improved multi-scale representation learning.

The stem in the decoder employs a similar architecture to take the features from the last layer of the decoder as the input features $\boldsymbol{X}_{\textrm{in}}'\in\mathbb{R}^{C\times\frac{H}{4}\times\frac{W}{4}\times\frac{D}{4}}$. This stem includes two convolutional linear projection layers, and each utilizes a $3\times3\times3$ convolutional layer, and a transposed convolutional layer with the stride of 2. A Batch Normalization and a GELU activation function are employed after each convolutional layer. To enhance final segmentation results with richer low-level features, the output features of FCSB $\hat{\boldsymbol{X}}_1$ and $\hat{\boldsymbol{X}}_2$ are added to this stem in the decoder. Thus, two linear projection layers generate features $\boldsymbol{X}_2'\in\mathbb{R}^{\frac{C}{2}\times\frac{H}{2}\times\frac{W}{2}\times\frac{D}{2}}$ and $\boldsymbol{X}_1'\in\mathbb{R}^{\frac{C}{2}\times H\times W\times D}$ as
\begin{align}
    \boldsymbol{X}_{\textrm{in}}'&=\boldsymbol{X}_{\textrm{in}}'+\hat{\boldsymbol{X}}_2,\\
    \boldsymbol{X}_2'&=\textrm{GELU}(\textrm{BN}(\textrm{Conv}_{3\times3\times3}(\boldsymbol{X}_{\textrm{in}}'))), \\
    \boldsymbol{X}_2'&=\textrm{GELU}(\textrm{BN}(\textrm{TransposedConv}_{3\times3\times3}(\boldsymbol{X}_2'))), \\
    \boldsymbol{X}_2'&=\boldsymbol{X}_2'+\hat{\boldsymbol{X}}_1,\\
    \boldsymbol{X}_1'&=\textrm{GELU}(\textrm{BN}(\textrm{Conv}_{3\times3\times3}(\boldsymbol{X}_2'))), \\
    \boldsymbol{X}_1'&=\textrm{GELU}(\textrm{BN}(\textrm{TransposedConv}_{3\times3\times3}(\boldsymbol{X}_1'))).
\end{align}

\section{Experiments and Results}
\subsection{Datasets}
We implemented the experiments on four segmentation tasks from four different datasets to evaluate the superiority of FEFormer and underline its potential to generalize across different segmentation tasks. These tasks differ in image modalities, segmentation complexity, number of structures to be segmented, and spatial and phenotypic heterogeneity (Table \ref{tab1}).

\textbf{Abdominal Multi-organ Segmentation.} We implemented abdominal multi-organ segmentation on the MICCAI 2022 AMOS Challenge dataset \citep{ji2022amos}. It consists of 300 abdominal CT images with voxel-level annotations of 15 organs (Spleen, Right kidney, Left kidney, Gall bladder, Esophagus, Liver, Stomach, Arota, Postcava, Pancreas, Right Adrenal Gland, Left Adrenal Gland, Duodenum, Bladder, and Prostate). Each CT volume consists of $67\thicksim369$ slices of $512\times512$ pixels with a slice spacing of $1.25\thicksim5.00$ mm.

\textbf{Hepatic Vessel Tumor Segmentation.} The Hepatic Vessel Tumor segmentation dataset is from the Medical Segmentation Decathlon (MSD) Challenge \citep{antonelli2022medical}. It consists of 303 CT scans with manual annotations. The target segmentation regions are the hepatic vessels (Vessel) and tumors within the liver (Tumor). They are obtained from patients with a variety of primary and metastatic liver tumors.

\textbf{Brain Tumor Segmentation.} The Multimodal Brain Tumor Segmentation dataset is from the MSD Challenge \citep{antonelli2022medical}. It comprises 484 multi-parametric Magnetic Resonance Imaging (MRI) scans with segmentation labels. Four modalities are available for each participant: Native T1-weighted image (T1w), post-contrast T1-weighted (T1Gd), T2-weighted (T2w), and T2 Fluid Attenuated Inversion Recovery (T2-FLAIR). Each subject has three foreground annotations: Edema (ED), Enhancing Tumor (ET), and Non-Enhancing Tumor (NET).

\textbf{Abdomen CT Organ Segmentation.}  The fourth dataset was from The Fast and Low GPU memory Abdominal oRgan sEgmentation (FLARE) challenge \citep{ma2022fast}. It consists of 361 CT images with voxel-wise annotations of four organs, including the liver, the kidneys, the spleen, and the pancreas. This dataset demonstrates a large diversity across various centers, vendors, phases, and diseases.

\begin{table}[!t]
\centering
\caption{Detailed information of the segmentation datasets used for evaluation, including segmentation tasks, imaging modalities, data size, and target segmentation class number.}
\label{tab1}
\resizebox{0.48\textwidth}{!}{
\begin{tabular}{c|c|c|c|c}
\toprule
Datasets & AMOS 2022 & Hepatic Vessel & Brain Tumor & FLARE \\
\midrule
Tasks & Multi-Organs & Tubular$\&$Tumor & Tissues & Abdominal Organs \\
Modalities   & CT    & CT  & Multi-modal MR & CT \\
Data size    & 300   & 303 & 484 & 361 \\
Class Number & 15    & 2   & 3   & 4   \\
\bottomrule
\end{tabular}}
\end{table}

\subsection{Implementation details}
The experiments were implemented using PyTorch. Models were trained for 1000 epochs with a batch size of 2 on NVIDIA Tesla A100 PCI-E Passive Single GPU with 40GB of GDDR5 memory. We used a joint loss function ($\mathcal{L}_{Seg}$) that consists of cross-entropy loss ($\mathcal{L}_{CE}$) and dice loss ($\mathcal{L}_{Dice}$) as
\begin{align}
    \mathcal{L}_{Seg}=\mathcal{L}_{CE}(\hat{y},y)+\mathcal{L}_{Dice}(\hat{y},y),
\end{align}
where $y$ and $\hat{y}$ denote the ground truth and predictions. The AdamW was used as the optimizer. The initial learning rate was set to $1e-3$, and decayed with a polynomial learning rate of $3e-5$ for multi-organ segmentation and hepatic vessel segmentation. Additionally, the initial learning rate was $5e-5$, and decayed with a rate of $3e-6$ in brain tumor segmentation.

We applied the similar strategy to CT and MRI scans from four datasets. The image intensities were clipped at the 5th and 95th percentiles, and then z-score normalization was applied to each volume. Subsequently, normalized scans were cropped to sub-volumes with the dimension of $96\times96\times96$ as input patches for model training. Data augmentation techniques were further implemented to improve model robustness. To be specific, patches were rotated between $[-30, 30]$ along three axes and also scaled between $(0.7,1.4)$ both with a probability of $0.2$. Subsequently, patches were mirrored along three axes with a probability of $0.5$. Zero-centered additive Gaussian noise with variance drawn from the distribution $U(0, 0.1)$, and brightness adjustments were added to each sample voxel with a probability of $0.15$ for each.

We employed the 5-fold cross-validation to generate reliable evaluation results on all datasets. Segmentation performance was evaluated by Dice Similarity Coefficient (DSC) and $95\%$ Hausdorff distance (HD95). The architecture complexity was evaluated by the number of parameters (Params), and the computational complexity was evaluated by the number of Floating Point Operations (FLOPs). The Wilcoxon signed-rank test was implemented to statistically quantify the differences in segmentation performance between FEFormer and SOTA methods.

\begin{table*}[!t]
\centering
\caption{Comparison of segmentation performance among FEFormer and SOTA methods on the 2022 AMOS Abdominal Multi-organ segmentation task. The performance was evaluated using the DSC and HD95 (Mean $\pm$ Standard Deviation). \textbf{Bold} represents the best results, and \underline{underline} represents the second best results. ($^*$: $p<0.01$ with Wilcoxon signed-rank test between FEFormer and SOTA methods.)}
\label{tab2}
\resizebox{\textwidth}{!}{
\begin{tabular}{c|c|c|c|c|c|c|c|c|c|c|c|c|c|c}
\toprule
Tasks &  VNet & Att UNet & nnU-Net & nnFormer & SegFormer & TransBTS & UNETR & Swin UNETR & UX-Net & MedNext & TransHRNet & VSmTrans & MixUNETR & FEFormer \\
\midrule
Spleen       & 95.18$\pm$8.58  & 96.14$\pm$8.39  & \underline{96.61}$\pm$6.18  & 95.27$\pm$8.43  & 92.03$\pm$10.79 & 95.91$\pm$8.30  & 93.68$\pm$10.14 
             & 95.93$\pm$8.31  & 95.95$\pm$8.80  & 95.88$\pm$8.35  & 95.83$\pm$8.62  & 96.00$\pm$8.31  & 95.72$\pm$8.70  & \textbf{97.74}$\pm$5.11 \\
R. kidney    & 95.05$\pm$8.04  & 95.90$\pm$5.82  & \underline{96.07}$\pm$5.74  & 94.04$\pm$6.49  & 92.74$\pm$6.59  & 95.59$\pm$5.85  & 92.83$\pm$11.48 
             & 95.74$\pm$5.98  & 95.73$\pm$6.22  & 95.73$\pm$3.05  & 95.54$\pm$7.11  & 95.34$\pm$8.11  & 95.24$\pm$7.51  & \textbf{96.92}$\pm$2.14 \\
L. kidney    & 95.05$\pm$8.60  & \underline{95.76}$\pm$7.37  & 95.53$\pm$9.47  & 94.00$\pm$9.25  & 91.88$\pm$11.28 & 94.85$\pm$10.68 & 92.78$\pm$12.39 
             & 95.10$\pm$9.78  & 95.09$\pm$10.11 & 94.91$\pm$8.82  & 95.02$\pm$9.98  & 95.03$\pm$10.07 & 94.99$\pm$9.33  & \textbf{96.71}$\pm$6.32 \\
Gall bladder & 77.81$\pm$25.70 & 81.86$\pm$24.20 & \underline{83.39}$\pm$22.82 & 82.57$\pm$20.41 & 71.07$\pm$25.59 & 81.91$\pm$21.85 & 71.10$\pm$27.63 
             & 80.62$\pm$24.91 & 80.97$\pm$24.53 & 80.85$\pm$24.12 & 81.25$\pm$24.06 & 82.04$\pm$23.57 & 81.40$\pm$23.67 & \textbf{85.18}$\pm$19.22 \\
Esophagus    & 81.04$\pm$10.25 & 83.71$\pm$9.36  & \underline{84.71}$\pm$9.65  & 79.52$\pm$11.68 & 69.97$\pm$12.90 & 82.77$\pm$10.03 & 77.57$\pm$12.75 
             & 83.25$\pm$9.76  & 83.52$\pm$9.92  & 82.47$\pm$10.32 & 82.97$\pm$9.28  & 83.82$\pm$8.93  & 82.97$\pm$9.28  & \textbf{86.40}$\pm$8.71 \\
Liver        & 96.83$\pm$2.29  & 97.45$\pm$1.99  & \underline{97.55}$\pm$1.85  & 96.83$\pm$2.47  & 95.56$\pm$3.31  & 97.29$\pm$1.86  & 95.71$\pm$3.93  
             & 97.21$\pm$2.34  & 97.21$\pm$2.95  & 97.22$\pm$2.33  & 97.24$\pm$2.67  & 97.38$\pm$2.00  & 97.20$\pm$2.55  & \textbf{97.80}$\pm$1.88 \\
Stomach      & 88.10$\pm$16.07 & 90.56$\pm$15.60 & \underline{91.31}$\pm$15.11 & 89.42$\pm$15.16 & 83.44$\pm$16.75 & 89.56$\pm$15.84 & 82.44$\pm$17.81 
             & 88.83$\pm$16.51 & 89.04$\pm$16.57 & 89.36$\pm$16.12 & 89.82$\pm$15.98 & 90.16$\pm$15.85 & 89.19$\pm$16.54 & \textbf{92.77}$\pm$15.10 \\
Arota        & 92.17$\pm$5.97  & 93.81$\pm$5.26  & 94.48$\pm$4.38  & 90.98$\pm$7.83  & 90.67$\pm$3.98  & 92.86$\pm$6.00  & 91.19$\pm$6.14  
             & 93.82$\pm$5.42  & 93.87$\pm$5.26  & 91.53$\pm$6.85  & 94.97$\pm$3.35  & \underline{95.07}$\pm$3.12  & 94.93$\pm$3.40  & \textbf{96.24}$\pm$2.35 \\
Postcava     & 86.64$\pm$7.77  & 89.41$\pm$6.72  & \underline{90.34}$\pm$5.82  & 83.48$\pm$13.37 & 83.58$\pm$6.78  & 87.92$\pm$7.85  & 84.18$\pm$7.94  
             & 89.18$\pm$6.42  & 89.22$\pm$6.88  & 87.18$\pm$8.93  & 90.11$\pm$5.64  & 90.29$\pm$5.61  & 89.92$\pm$5.76  & \textbf{92.82}$\pm$5.44 \\
Pancreas     & 81.38$\pm$12.63 & 84.50$\pm$11.95 & \underline{85.41}$\pm$11.76 & 79.24$\pm$14.72 & 75.66$\pm$13.43 & 82.63$\pm$13.54 & 76.91$\pm$15.31 
             & 83.07$\pm$13.07 & 83.67$\pm$13.03 & 83.18$\pm$13.03 & 84.04$\pm$12.20 & 84.61$\pm$11.84 & 83.87$\pm$12.77 & \textbf{88.18}$\pm$9.08 \\
R.A. gland   & 73.39$\pm$13.23 & 75.12$\pm$14.65 & \underline{76.53}$\pm$13.36 & 69.93$\pm$14.13 & 59.44$\pm$11.31 & 73.43$\pm$14.71 & 70.87$\pm$14.73 
             & 75.61$\pm$13.28 & 75.44$\pm$14.16 & 74.89$\pm$13.73 & 75.44$\pm$12.31 & 75.89$\pm$12.04 & 75.38$\pm$12.52 & \textbf{79.32}$\pm$8.82 \\
L.A. gland   & 74.51$\pm$14.03 & 75.55$\pm$14.95 & \underline{77.23}$\pm$13.85 & 70.65$\pm$14.31 & 55.82$\pm$13.23 & 74.37$\pm$15.02 & 68.80$\pm$17.55 
             & 75.82$\pm$14.72 & 75.89$\pm$15.03 & 75.15$\pm$14.79 & 76.12$\pm$14.45 & 76.38$\pm$14.37 & 76.04$\pm$14.08 & \textbf{81.14}$\pm$8.95 \\
Duodenum     & 75.39$\pm$14.89 & 79.21$\pm$15.00 & \underline{80.79}$\pm$14.68 & 73.08$\pm$16.76 & 68.72$\pm$14.41 & 77.89$\pm$15.21 & 69.04$\pm$15.57 
             & 77.96$\pm$14.78 & 77.83$\pm$15.53 & 77.47$\pm$15.74 & 79.11$\pm$14.52 & 80.18$\pm$14.10 & 78.60$\pm$14.68 & \textbf{83.16}$\pm$10.18 \\
Bladder      & 85.04$\pm$17.50 & 88.07$\pm$15.38 & \underline{89.37}$\pm$13.74 & 86.71$\pm$14.47 & 78.99$\pm$18.00 & 87.95$\pm$14.65 & 79.10$\pm$21.24 
             & 86.92$\pm$16.32 & 86.91$\pm$18.20 & 87.18$\pm$15.56 & 87.60$\pm$15.36 & 87.71$\pm$15.12 & 87.53$\pm$15.41 & \textbf{90.70}$\pm$12.06 \\
Prostate     & 79.48$\pm$19.51 & 82.96$\pm$19.33 & \underline{83.72}$\pm$20.23 & 81.85$\pm$17.88 & 75.16$\pm$18.84 & 82.30$\pm$18.86 & 75.90$\pm$20.78 
             & 80.92$\pm$20.15 & 81.56$\pm$20.79 & 80.72$\pm$20.64 & 81.48$\pm$19.84 & 81.29$\pm$20.73 & 81.41$\pm$19.57 & \textbf{86.56}$\pm$18.12 \\
\midrule
Mean DSC     & 85.15$\pm$15.76 & 87.34$\pm$14.97 & \underline{88.21}$\pm$14.31 & 84.51$\pm$15.77 & 78.99$\pm$18.25 & 86.49$\pm$15.13 & 81.49$\pm$18.15  
             & 86.68$\pm$15.34 & 86.81$\pm$15.65 & 86.26$\pm$15.38 & 87.10$\pm$14.88 & 87.42$\pm$14.76 & 86.97$\pm$14.92 & \textbf{90.11}$^*\pm$10.60 \\
Mean HD95    & 2.63$\pm$3.51   & \underline{1.93}$\pm$2.59   & 2.02$\pm$2.69   & 2.73$\pm$3.11   & 3.32$\pm$2.79   & 2.15$\pm$2.38   & 3.56$\pm$3.54 
             & 2.38$\pm$3.24   & 2.36$\pm$3.38   & 2.18$\pm$2.42   & 2.04$\pm$2.70   & 1.96$\pm$2.24   & 2.17$\pm$2.95   & \textbf{1.78}$^*\pm$2.04 \\
\bottomrule
\end{tabular}}
\end{table*}

\begin{table*}[!t]
\centering
\caption{Comparison of segmentation performance among FEFormer and SOTA methods on the Hepatic Vessel Tumor segmentation and Abdomen Organ segmentation tasks. The performance was evaluated using the DSC and HD95 (Mean $\pm$ Standard Deviation). \textbf{Bold} represents the best results, and \underline{underline} represents the second best results. ($^*$: $p<0.01$ with Wilcoxon signed-rank test between FEFormer and SOTA methods.)}
\label{tab3}
\resizebox{\textwidth}{!}{
\begin{tabular}{c|cc|cc|cc|cccc}
\toprule
Tasks & \multicolumn{4}{c|}{Hepatic Vessel Tumor Segmentation} & \multicolumn{6}{c}{FLARE Abdomen Organ segmentation} \\
\midrule
Methods    &  Mean DSC & Mean HD95 & Vessel DSC & Tumor DSC & Mean DSC & Mean HD95 & Liver DSC & Kidney DSC & Spleen DSC & Pancreas DSC \\
\midrule
V-Net      & 64.92$\pm$21.41 & 10.98$\pm$9.46  & 62.59$\pm$12.57 & 67.25$\pm$27.34 
           & 93.69$\pm$8.34  & 1.81$\pm$1.33 & 98.26$\pm$1.26  & 96.18$\pm$3.37  & 98.00$\pm$2.17  & 82.34$\pm$9.25 \\
Att U-Net  & 65.14$\pm$21.17 & 11.35$\pm$12.03 & 61.80$\pm$12.75 & 68.48$\pm$26.67
           & 94.32$\pm$7.64  & \underline{1.53}$\pm$0.99 & \underline{98.55}$\pm$0.80  & 96.55$\pm$3.42  & \underline{98.28}$\pm$1.00  & 83.91$\pm$8.56 \\
nnU-Net    & 65.96$\pm$20.94 & 11.18$\pm$11.92 & 62.73$\pm$12.71 & 69.20$\pm$26.36 
           & 94.30$\pm$7.68  & 1.54$\pm$1.03 & 98.54$\pm$0.83  & \underline{96.62}$\pm$3.11  & 98.27$\pm$1.02  & 83.76$\pm$8.59 \\
nnFormer   & \underline{66.26}$\pm$20.55 & \underline{10.47}$\pm$9.12  & \underline{63.15}$\pm$12.05 & 69.36$\pm$26.09
           & 93.95$\pm$7.52  & 1.95$\pm$3.12 & 98.32$\pm$1.05  & 96.08$\pm$2.05  & 98.02$\pm$1.18  & 83.38$\pm$8.22 \\
SegFormer  & 59.11$\pm$21.54 & 13.29$\pm$13.44 & 54.36$\pm$10.42 & 63.87$\pm$27.82
           & 92.77$\pm$8.63  & 2.22$\pm$1.41 & 97.93$\pm$1.05  & 95.37$\pm$2.90  & 97.31$\pm$1.42  & 80.45$\pm$9.03 \\
TransBTS   & 64.62$\pm$21.33 & 12.00$\pm$12.04 & 61.39$\pm$12.70 & 67.84$\pm$26.98
           & 94.05$\pm$8.11  & 1.60$\pm$1.14 & 98.47$\pm$0.99  & 96.46$\pm$3.50  & 98.26$\pm$1.10  & 83.00$\pm$9.14 \\
UNETR      & 58.57$\pm$24.21 & 13.76$\pm$15.35 & 62.39$\pm$12.37 & 54.75$\pm$31.47
           & 92.54$\pm$10.22 & 2.27$\pm$1.91 & 97.97$\pm$1.71  & 96.08$\pm$3.21  & 97.19$\pm$6.28  & 78.93$\pm$10.77 \\
Swin UNETR & 62.66$\pm$22.52 & 12.75$\pm$12.76 & 61.99$\pm$12.64 & 63.33$\pm$29.23
           & 93.80$\pm$8.57  & 1.70$\pm$1.31 & 98.37$\pm$1.57  & 96.51$\pm$3.23  & 98.01$\pm$3.57  & 82.33$\pm$9.52 \\
UX Net     & 63.11$\pm$22.21 & 11.95$\pm$11.52 & 62.30$\pm$12.41 & 63.92$\pm$28.82
           & 93.92$\pm$8.43  & 1.69$\pm$1.51 & 98.42$\pm$1.34  & 96.50$\pm$3.45  & 98.21$\pm$1.74  & 82.59$\pm$9.68 \\
MedNeXt    & 66.11$\pm$20.55 & 10.69$\pm$9.05  & 62.57$\pm$12.41 & \underline{69.64}$\pm$25.80
           & 94.19$\pm$7.82  & 1.61$\pm$1.18 & 98.45$\pm$0.95  & 96.52$\pm$3.38  & 98.26$\pm$0.97  & 83.53$\pm$8.80 \\
TransHRNet & 65.22$\pm$21.04 & 11.28$\pm$11.96 & 61.92$\pm$12.66 & 68.51$\pm$26.55
           & 94.31$\pm$7.91  & 1.55$\pm$1.10 & 98.42$\pm$1.66  & 96.48$\pm$3.34  & 98.10$\pm$1.21  & \underline{84.24}$\pm$7.80 \\
VSmTrans   & 64.93$\pm$20.46 & 13.89$\pm$19.34 & 62.25$\pm$12.49 & 67.60$\pm$25.83
           & \underline{94.34}$\pm$7.33  & 1.57$\pm$1.09 & 98.46$\pm$0.97  & 96.55$\pm$3.25  & 98.25$\pm$1.13  & 84.10$\pm$7.74 \\
MixUNETR   & 63.61$\pm$21.28 & 12.42$\pm$14.14 & 62.19$\pm$12.74 & 65.03$\pm$27.18
           & 94.09$\pm$7.83  & 1.63$\pm$1.16 & 98.40$\pm$1.25  & 96.55$\pm$3.26  & 98.13$\pm$2.35  & 83.30$\pm$8.38 \\          
\midrule
FEFormer   & \textbf{67.97}$^*\pm$20.08 & \textbf{9.94}$^*\pm$8.98 & \textbf{64.96}$\pm$10.32 & \textbf{70.98}$\pm$25.66 & \textbf{95.02}$^*\pm$5.96 
           & \textbf{1.40}$^*\pm$1.05& \textbf{98.65}$\pm$0.90 & \textbf{97.25}$\pm$1.27 & \textbf{98.42}$\pm$0.92 & \textbf{85.74}$\pm$5.01 \\
\bottomrule
\end{tabular}}
\end{table*}

\begin{table*}[!t]
\centering
\caption{Comparison of segmentation performance among FEFormer and SOTA methods on the Brain Tumor segmentation task. The performance was evaluated using the DSC (Mean $\pm$ Standard Deviation) and HD95 (Mean $\pm$ Standard Deviation, failure rate with HD95$>$100mm). \textbf{Bold} represents the best results, and \underline{underline} represents the second best results. ($^*$: $p<0.01$ with Wilcoxon signed-rank test between FEFormer and SOTA methods.)}
\label{tab4}
\resizebox{0.8\textwidth}{!}{
\begin{tabular}{c|cc|ccc}
\toprule
Methods    & Mean DSC & Mean HD95 & ET DSC & ED DSC & NET DSC \\
\midrule
V-Net      & 73.03$\pm$21.52 & 5.98$\pm$9.76, 0.207$\%$  & 77.61$\pm$22.00 & 79.80$\pm$12.81 & 61.73$\pm$23.40 \\
Att U-Net  & 73.53$\pm$21.38 & 5.62$\pm$9.58, 0.000$\%$  & 78.83$\pm$21.44 & 80.45$\pm$11.98 & 61.37$\pm$23.35 \\
nnU-Net    & 73.47$\pm$21.34 & 5.70$\pm$9.59, 0.000$\%$  & 78.78$\pm$21.55 & 80.35$\pm$11.87 & 61.35$\pm$23.21 \\
nnFormer   & 73.62$\pm$21.32 & 5.42$\pm$9.06, 0.000$\%$  & 79.03$\pm$20.86 & 80.47$\pm$11.74 & 61.36$\pm$23.18\\
SegFormer  & 44.77$\pm$26.47 & 14.42$\pm$9.63, 0.413$\%$ & 37.73$\pm$26.49 & 63.06$\pm$17.16 & 33.41$\pm$24.38 \\
TransBTS   & 73.75$\pm$21.20 & 5.63$\pm$9.86, 0.000$\%$  & 78.91$\pm$21.58 & \underline{80.78}$\pm$11.59 & 61.64$\pm$22.92 \\
UNETR      & 72.59$\pm$21.65 & 5.85$\pm$9.39, 0.000$\%$  & 78.00$\pm$21.78 & 79.83$\pm$11.85 & 60.00$\pm$23.47 \\
Swin UNETR & 73.76$\pm$21.13 & 5.37$\pm$8.02, 0.000$\%$  & 79.12$\pm$20.96 & 80.58$\pm$11.58 & 61.66$\pm$23.35 \\
UX Net     & 58.23$\pm$24.23 & 11.07$\pm$9.27, 0.000$\%$ & 60.11$\pm$26.20 & 68.79$\pm$14.47 & 45.81$\pm$24.42 \\
MedNeXt    & 73.75$\pm$21.39 & 5.50$\pm$9.29, 0.000$\%$  & 78.81$\pm$21.77 & 80.72$\pm$11.94 & 61.77$\pm$23.21 \\
TransHRNet  & 73.71$\pm$21.30 & 5.32$\pm$8.99, 0.000$\%$  & 78.58$\pm$21.86 & 80.70$\pm$11.82 & \underline{61.86}$\pm$23.16 \\
VSmTrans   & \underline{73.87}$\pm$21.28 & \underline{5.31}$\pm$9.14, 0.000$\%$  & \underline{79.18}$\pm$21.33 & 80.69$\pm$11.95 & 61.80$\pm$23.25 \\
MixUNETR   & 73.68$\pm$21.52 & 5.34$\pm$8.79, 0.207$\%$  & 78.52$\pm$22.34 & 80.71$\pm$11.79 & 61.83$\pm$23.21 \\
\midrule
FEFormer   & \textbf{74.97}$^*\pm$19.71 & \textbf{5.01}$^*\pm$7.76, 0.000$\%$ & \textbf{80.30}$\pm$20.54 & \textbf{81.72}$\pm$9.66 & \textbf{62.89}$\pm$21.65 \\
\bottomrule
\end{tabular}}
\end{table*}

\begin{table}[!t]
\centering
\caption{Comparison of model complexity among FEFormer and SOTA methods. Params and FLOPs were evaluated using input patches with dimensions of $96\times96\times96$.}
\label{tab5}
\begin{tabular}{c|cc}
\toprule
Methods    & Params (M) & FLOPs (G) \\
\midrule
VNet       & 45.66   & 370.52 \\
Att U-Net  & 69.08   & 360.98 \\
nnU-Net    & 68.38   & 357.13 \\
nnFormer   & 149.33  & 284.28 \\
SegFormer  & 4.50    & 5.02   \\
TransBTS   & 31.58   & 110.69 \\
UNETR      & 92.78   & 82.73  \\
Swin UNETR & 62.19   & 329.28 \\
UX Net     & 53.01   & 632.33 \\
MedNeXt    & 11.65   & 178.05 \\
TransHRNet & 36.86   & 340.33 \\
VSmTrans   & 50.39   & 358.21 \\
MixUNETR   & 62.03   & 329.99 \\
\midrule
FEFormer   & 18.54   & 39.13  \\
\bottomrule
\end{tabular}
\end{table}

\subsection{Comparison with State-of-the-arts}
We compared the performance of FEFormer with various recent 3D state-of-the-art (SOTA) segmentation models. These methods include
\begin{itemize}
    \item CNN-based methods: VNet \citep{milletari2016v}, nnU-Net \citep{isensee2021nnu}, Attention gated U-Net (Att UNet) \citep{oktay2018attention}
    \item ViT-based methods: nnFormer \citep{zhou2023nnformer}, SegFormer \citep{perera2024segformer3d}
    \item Transformer-like CNN-based methods: 3D UX Net \citep{lee20223d} and MedNext \citep{roy2023mednext}
    \item Hybrid CNN-ViT-based methods: TransBTS \citep{wang2021transbts}, UNETR \citep{hatamizadeh2022unetr}, Swin UNETR \citep{hatamizadeh2021swin}, TransHRNet \citep{yan20233d}, VSmTrans \citep{liu2024vsmtrans}, and MixUNETR \citep{shen2025mixunetr}
\end{itemize}

\textbf{AMOS Multi-organ segmentation.} FEFormer outperformed all compared SOTA methods on this AMOS abdominal multi-organ segmentation task, achieving the best mean DSC of $90.11\%$ and the lowest mean HD95 of $1.78$ mm, with statistical significance ($p < 0.01$) (Table~\ref{tab2}). Additionally, FEFormer achieved the highest DSC across all 15 organs, demonstrating its strong capability in accurately segmenting both large anatomical structures (e.g., liver and spleen) and challenging small or low-contrast organs (e.g., pancreas, adrenal glands, and duodenum). Specifically, FEFormer achieved superior performance compared with CNN-based methods such as VNet, Attention U-Net, and nnU-Net. Particularly, it outperformed the second best method, nnU-Net, by $1.9$ DSC points and $0.24$ HD95 points. Compared with transformer-based models such as nnFormer and SegFormer, FEFormer achieved notably higher accuracy, especially on challenging organs such as the gallbladder ($85.18\%$ vs. $82.57\%$ of nnFormer) and pancreas ($88.18\%$ vs. $79.24\%$ of nnFormer), highlighting its advantage in capturing fine structural details and segmenting largely varying-sized organs. Furthermore, FEFormer consistently achieved higher DSC scores and lower HD95 points than all other hybrid methods. Particularly, FEFormer outperformed the best hybrid method, VSmTrans, by $2.69$ points in mean DSC and $0.18$ points in mean HD95.

In addition to segmentation accuracy, FEFormer demonstrated a favorable trade-off between performance and computational efficiency (Table~\ref{tab5}). Despite achieving the best overall accuracy, FEFormer required only $18.54$M parameters and $39.13$G FLOPs, substantially lower than the best CNN-based method (nnU-Net, 68.38M Params, 357.13G FLOPs), the best ViT-based method (nnFormer, 149.33M parameters, 284.28G FLOPs), and the best hybrid method (VSmTrans, 50.39M parameters, 358.21G FLOPs). Moreover, when compared with lightweight models such as SegFormer, FEFormer achieved significantly higher segmentation accuracy by $11.12$ points in mean DSC and $1.54$ points in mean HD95 with a moderate increase in complexity. These results demonstrated that FEFormer effectively balanced global contextual modeling and local detail preservation while maintaining high segmentation efficiency.

\textbf{Hepatic Vessel Tumor segmentation.} In the hepatic vessel tumor segmentation task, FEFormer achieved the best overall performance than SOTA methods, achieving a mean DSC of $67.97\%$ and a mean HD95 of $9.94$ mm, with statistical significance ($p < 0.01$) (Table~\ref{tab3}). Specifically, FEFormer improved the mean DSC by approximately $2.0\%$ and $1.7\%$ than two strong baselines, nnU-Net and nnFormer respectively. Additionally, FEFormer outperformed the best two hybrid method MedNeXt and TransHRNet by $1.86$ and $2.75$ points in mean DSC and $0.75$ and $1.34$ points in mean HD95. At the class level, FEFormer demonstrated the highest DSC for both vessel ($64.96\%$) and tumor ($70.98\%$), reflecting its ability to simultaneously segment thin, low-contrast vascular structures and highly heterogeneous tumor regions. Specifically, vessel segmentation is inherently challenging due to its complex topology, small caliber, and discontinuities; however, FEFormer showed superior performance compared to other methods. For tumor segmentation, which often suffered from irregular shapes and ambiguous boundaries, FEFormer achieved superior accuracy, indicating enhanced capability in capturing intra-tumoral heterogeneity and boundary details. Therefore, these results demonstrated that FEFormer was effective for challenging medical segmentation tasks involving complex geometry and high variability, delivering more accurate, robust, and reliable predictions.

\textbf{FLARE Abdomen Organ segmentation.} FEFormer achieved superior performance than other SOTA methods in the FLARE abdomen organ segmentation task, achieving $95.02$ in mean DSC and $1.40$ in mean HD95 with statistical significance (Table~\ref{tab3}). FEFormer outperformed recent methods, such as MedNeXt, VSmTrans and, MixUNETR significantly. Additionally, FEFormer consistently outperformed SOTA methods across all evaluated organs, achieving the best performance on all structures, including liver ($98.65\%$), kidney ($97.25\%$), spleen ($98.42\%$), and pancreas ($85.74\%$). Specifically, it outperformed the second best method VSmTrans by $0.19$, $0.70$, $0.17$, and $1.64$ higher DSC points in these four organs.

\textbf{Brain Tumor segmentation.} FEFormer achieved the best overall performance on the brain tumor segmentation task than all SOTA methods by achieving the highest mean DSC of $74.97\%$ and the lowest mean HD95 of $5.01$ mm with statistical significance ($p < 0.01$) (Table~\ref{tab4}). Specifically, FEFormer demonstrated superior performance than nnFormer and Swin UNETR which were initially proposed for brain tumor segmentation by $1.35$ and $1.38$ points in mean DSC and $0.41$ and $0.38$ points in mean HD95. Additionally, FEFormer improved the mean DSC and mean HD95 by more than $1.50$ and $0.69$ points over nnU-Net, and outperformed the second best model VSmTrans by $1.10$ points and $0.30$ points in mean DSC and mean HD95, respectively. At the sub-region level, FEFormer consistently demonstrated the highest DSC scores for all tumor regions, including ET, ED, and NET. Notably, improvements were more significant in the challenging ET and NET regions, which typically exhibited high variability and low contrast. Additionally, FEFormer maintained a zero failure rate (HD95 $> 100$ mm), further demonstrating its stability and reliability.

\begin{figure*}[!t]
\centering
\includegraphics[width=\textwidth]{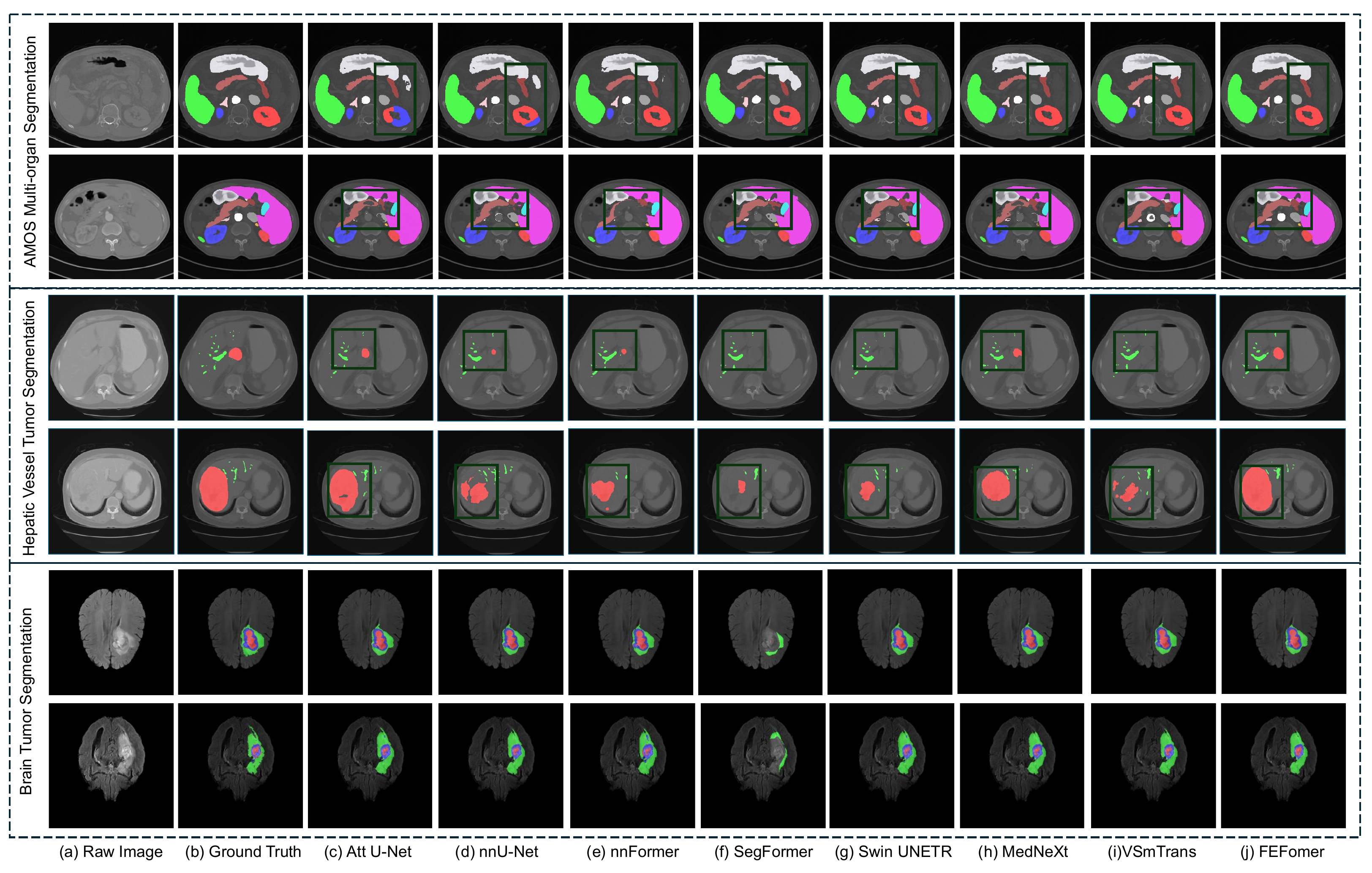}
\caption{Qualitative comparison between FEFormer and (c) Att UNet, (d) nnU-Net, (e) nnFormer, (f) SegFormer, (g) Swin UNETR, (h) MedNext, and (i) VSmTrans. Results are shown across three public datasets, including the AMOS Abdominal Multi-organ dataset, the Hepatic Vessel Tumor dataset, and the Brain Tumor dataset. Red boxes mark the regions where FEFormer demonstrates better segmentation results than other methods.}
\label{fig4}
\end{figure*}

\subsection{Visual assessment of segmentation results}
To further demonstrate the superior performance of FEFormer over SOTA methods, we conducted a qualitative visual assessment of segmentation results (Figure \ref{fig4}). The results of this visual analysis demonstrated that the proposed frequency-enhanced design enabled FEFormer to jointly model global context and fine structural details by overcoming key limitations of both CNN- and ViT-based architectures, leading to improved segmentation quality across diverse and challenging scenarios.

FEFormer demonstrates a superior ability to capture and utilize global contextual information by employing frequency-domain attention mechanisms compared to CNN-based models, including Att U-Net, nnU-Net, and MedNeXt. This enables FEFormer to maintain better anatomical coherence and more effectively disambiguate locally ambiguous features when FEFormer segments anatomically complex structures with large variations in shape and size. Therefore, FEFormer acheived more accurate and consistent segmentation results for anatomically complex structures with large variations in shape and size, such as the pancreas in multi-organ segmentation, as well as hepatic tumors and brain lesions. In contrast, CNN-based methods often demonstrated inferior qualitative results in these complex anatomical structures due to their limited receptive field and insufficient modeling of global dependencies, leading to fragmented or inconsistent predictions in such challenging scenarios.

Additionally, FEFormer demonstrated superior performance and ability to preserve and leverage high-frequency information for guiding segmentation compared with ViT-based and hybrid methods, including nnFormer, SegFormer, Swin UNETR, and VSmTrans. By explicitly modeling high-frequency spectral components and enhancing high-frequency representations, FEFormer achieved more precise delineation of edges and boundaries, so it demonstrated superior qualitative segmentation results in boundary regions of anatomical structures than those ViT-based methods. Additionally, FEFormer achieved better segmentation results in small anatomical structures and tubular structures, such as hepatic vessels due to the combination of high-frequency and low-frequency information. FEFormer generated more continuous and anatomically plausible predictions, while ViT-based and hybrid methods generated segmentation results with discontinuities or mis-segmentation.

Moreover, FEFormer demonstrated superior capability in capturing fine-grained structural details, thus reducing label misassignment and semantic confusion between neighboring anatomical regions. Specifically, other SOTA models frequently misclassified neighboring structures by generating segmentation masks of one structure as belonging to another neighboring anatomical structure. In contrast, FEFormer generated more precise and well-separated segmentation masks, highlighting its robustness in complex anatomical contexts.

\subsection{Ablation study}
\subsubsection{The impact of different modules}
\begin{figure*}[!t]
\centering
\includegraphics[width=0.6\textwidth]{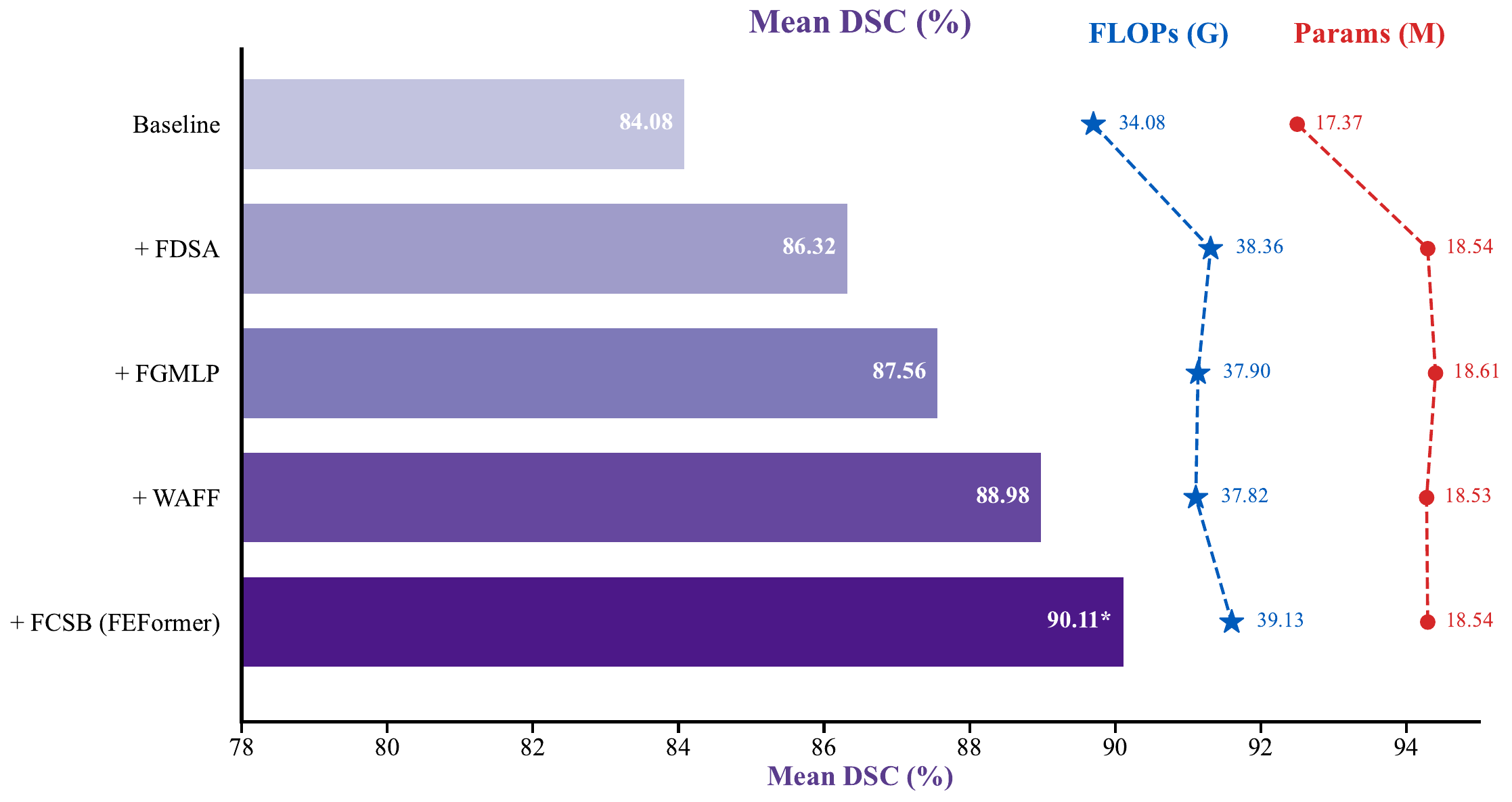}
\caption{Road-map visualization of the cumulative impact of the proposed modules on segmentation performance and model complexity in the 2022 AMOS multi-organ segmentation task. The progressive incorporation of the FDSA, FGMLP, WAFF, and FCSB into the plain ViT architecture consistently improved the mean DSC from $84.08\%$ to $90.11\%$. Meanwhile, the corresponding changes in computational cost demonstrated that the proposed FEFormer achieved substantial accuracy gains with only modest increases in model complexity.}
\label{fig5}
\end{figure*}

We conduct a comprehensive ablation study on the 2022 AMOS multi-organ segmentation task to investigate the separate and joint contributions of the proposed modules on segmentation performance and model complexity, including FDSA, FGMLP, WAFF, and FCSB, (Table~\ref{tab6} and Fig.~\ref{fig5}). The progressive performance improvements across different configurations highlighted that all modules contributed positively, and their combination yielded a synergistic effect, leading to substantial improvements in segmentation accuracy while maintaining comparable model complexity.

The plain ViT-based segmentation models with standard SA and MLP modules achieved a mean Dice score of $84.08\%$ and an HD95 of $2.86$. The incorporation of FDSA into plain ViT network to replace the standard SA module significantly improved the performance by $2.24$ points in DSC and $0.70$ points in HD95 with a slight increases in Params and FLOPs. It demonstrated the effectiveness of FDSA in capturing long-range dependencies and enhancing global contextual representation. Additionally, when FGMLP was incorporated into the plain ViT network to replace the standard MLP module, the model performance was improved by $1.13$ DSC points and $0.65$ HD95 points while slightly increasing model complexity. Thus, by disentangling and adaptively reweighting frequency components, FGMLP enhances both global structure representation and fine-grained detail preservation. Furthermore, when FDSA and FGMLP were jointly employed in the Frequency-enhanced Transformer block, the performance was enhanced to $87.56\%$ Dice and $2.10$ mm HD95 by $2.48$ DSC points and $0.91$ HD95 points with only $0.24$M Params and $3.82$G FLOPs increased, thus demonstrating the joint impact of these two modules on enhancing segmentation accuracy. Plain ViT network employed the concatenation operation to combine skip-connected encoder features and upsampled decoder features, but introducing WAFF to replace it enabled more effective and efficient feature fusion. Specifically, introducing WAFF further improved the performance by $1.42$ DSC points and $0.12$ HD95 points and decreased $0.07$ Params and $0.08$G FLOPs. Finally, incorporating FCSB generated the FEFormer and thus achieved the best performance of $90.11\%$ Dice and $1.78$ mm HD95 with statistical significance ($p<0.01$).

\begin{table}[t!]
\centering
\caption{Ablation study on the impact of the FDSA, FGMLP, WAFF, and FCSB modules in the 2022 AMOS multi-organ segmentation task. The performance was evaluated using the DSC and HD95 (Mean $\pm$ Standard Deviation). \textbf{Bold} represents the best results. The Params and FLOPs were evaluated using input patches with dimensions of $96\times96\times96$. ($^*$: $p<0.01$ with Wilcoxon signed-rank test between FEFormer and other designs.)}
\label{tab6}
\resizebox{0.48\textwidth}{!}{
\begin{tabular}{cccc|cc|cc}
\toprule
\multicolumn{4}{c|}{Modules} & \multicolumn{2}{c|}{Segmentation performance} & \multicolumn{2}{c}{Model complexity} \\
\midrule
FDSA & FGMLP & WAFF & FCSB & Mean DSC & Mean HD95 & Params (M) & FLOPs (G) \\
\midrule
           &            &            &            & 84.08$\pm$14.82 & 2.86$\pm$3.22 & 17.37 & 34.08 \\
\checkmark &            &            &            & 86.32$\pm$13.48 & 2.16$\pm$2.37 & 18.54 & 38.36 \\
           & \checkmark &            &            & 86.21$\pm$13.65 & 2.21$\pm$2.45 & 17.43 & 36.32 \\
\checkmark & \checkmark &            &            & 87.56$\pm$12.05 & 1.95$\pm$2.26 & 18.61 & 37.90 \\
\checkmark & \checkmark & \checkmark &            & 88.98$\pm$11.44 & 1.83$\pm$2.15 & 18.53 & 37.82 \\
\checkmark & \checkmark & \checkmark & \checkmark & \textbf{90.11}$^*\pm$10.60 & \textbf{1.78}$^*\pm$2.04 & 18.54 & 39.13 \\
\bottomrule
\end{tabular}}
\end{table}

\subsubsection{The effectiveness of FDSA and FGMLP}
To investigate the effectiveness of the proposed FDSA and FGMLP modules, we conducted ablation studies to evaluate the impact of their internal components on segmentation performance on the AMOS multi-organ segmentation task (Table \ref{tab7}).

For the FDSA module, replacing standard spatial-domain self-attention with frequency-domain self-attention improved the mean DSC from $87.68\%$ to $89.06\%$ and reduced the mean HD95 from $2.07$ mm to $1.85$ mm. Further incorporating the multi-frequency dynamic mechanism enhanced the performance by $1.05$ DSC points and $0.07$ HD95 points. These results demonstrated the effectiveness of the frequency-domain self-attention and multi-frequency dynamic mechanisms in generating more accurate segmentation results. For the FGMLP module, incorporating a gating mechanism into the standard MLP module improved the DSC by $0.74$ points to $88.75\%$ and HD95 by $0.12$ points to $1.90$ mm. Additionally, further incorporating the selective frequency decomposition mechanism enhanced the segmentation performance by $1.36$ DSC points and $0.12$ HD95 points. Thus, employing the gating and selective frequency decomposition mechanisms in the MLP module improved the segmentation performance of FEFormer.

\begin{table}[t!]
\centering
\caption{Ablation study on different mechanisms of the FDSA and FGMLP modules on the 2022 AMOS multi-organ segmentation task. The performance was evaluated using the DSC and HD95 (Mean $\pm$ Standard Deviation). \textbf{Bold} represents the best results. ($^*$: $p<0.01$ with Wilcoxon signed-rank test between FEFormer and other configurations.)}
\label{tab7}
\resizebox{0.48\textwidth}{!}{
\begin{tabular}{c|c|cc}
\toprule
Modules & Configurations & Mean DSC & Mean HD95  \\
\midrule
\multirow{3}{*}{\rotatebox[origin=c]{90}{FDSA}} 
& Standard self-attention             & 87.85$\pm$11.77 & 2.07$\pm$2.32 \\
& Frequency-domain self-attention     & 89.06$\pm$11.20 & 1.85$\pm$2.20 \\
& + Multi-frequency dynamic mechanism & \textbf{90.11}$^*\pm$10.60 & \textbf{1.78}$^*\pm$2.04 \\
\midrule
\multirow{3}{*}{\rotatebox[origin=c]{90}{FGMLP}} 
& Standard MLP                                  & 88.01$\pm$11.60 & 2.02$\pm$2.25 \\
& Gating mechanism                              & 88.75$\pm$11.48 & 1.90$\pm$2.18 \\
& + Selective frequency decomposition mechanism & \textbf{90.11}$^*\pm$10.60 & \textbf{1.78}$^*\pm$2.04 \\
\bottomrule
\end{tabular}}
\end{table}

\subsection{Comparing WAFF with other fusion modules}
To validate the superiority of WAFF, we compared it with two advanced feature fusion strategies: Attentional Feature Fusion (AFF) \citep{dai2021attentional}  and Dynamic Feature Fusion (DFF) \citep{yang2026d}. These methods employed channel-wise and spatial-wise attention mechanisms to fuse features adaptively and dynamically based on spatial-domain relationship. However, WAFF enabled the fusion of features based on their semantics, thus aligning their global structures and fine-grained details for more effective fusion. 

We incorporated WAFF and these two modules into the FEFormer by replacing simple concatenation, and compared their impacts on segmentation performance and model complexity (Table~\ref{tab7}). Specifically, AFF leveraged channel-wise attention to emphasize informative features during fusion, but incorporating WAFF outperformed it by $1.29$ DSC points and $0.11$ HD95 points with $0.09$M fewer Params and similar FLOPs. Additionally, DFF implemented input-adaptive feature fusion by utilizing channel-wise and spatial-wise dynamic mechanisms. However, employing WAFF improved by $0.54$ DSC points and $0.03$ HD95 points over DFF with $0.43$M fewer Params and $0.09$G fewer FLOPs. Thus, these results demonstrated that WAFF provided a more effective and robust feature fusion mechanism than conventional spatial-domain feature fusion approaches, contributing significantly to the performance improvements of FEFormer.

\begin{table}[!t]
\centering
\caption{Comparison of segmentation performance and model complexity between the WAFF module and the AFF and DFF modules on the 2022 AMOS multi-organ segmentation task. \textbf{Bold} represents the best results. The Params and FLOPs were evaluated using input patches with dimensions of $96\times96\times96$. ($^*$: $p<0.01$ with Wilcoxon signed-rank test between WAFF and AFF, DFF modules.)}
\resizebox{0.45\textwidth}{!}{
\begin{tabular}{c|cc|cc}
\toprule
Modules & Mean DSC & Mean HD95 & Params (M) & FLOPs (G)  \\
\midrule
AFF           & 88.82$\pm$12.02 & 1.89$\pm$2.21 & 18.63 & 39.13 \\
DFF           & 89.54$\pm$11.78 & 1.81$\pm$2.10 & 18.97 & 39.22 \\
WAFF          & \textbf{90.11}$^*\pm$10.60 & \textbf{1.78}$^*\pm$2.04 & 18.54 & 39.13 \\
\bottomrule
\end{tabular}}
\label{tab8}
\end{table}

\begin{table}[!t]
\centering
\caption{Comparison of segmentation performance between FEFormer and other SOTA methods on the internal and external evaluation of the FLARE organ segmentation task. \textbf{Bold} indicates the best performance. Segmentation performance was reported as DSC and HD95 (Mean $\pm$ Standard Deviation). ($^*$: $p<0.01$ with the Wilcoxon signed-rank test between FEFormer and SOTA methods.)}
\resizebox{0.48\textwidth}{!}{
\begin{tabular}{c|cc|cc|cc}
\toprule
\multirow{2}{*}{Methods} & \multicolumn{2}{c|}{FLARE (Internal)} & \multicolumn{2}{c|}{FLARE (External)} & \multicolumn{2}{c}{Generalization Gaps} \\
\cline{2-7}
 & Mean DSC & Mean HD95 & Mean DSC & Mean HD95 & Mean DSC & Mean HD95 \\
\midrule
V-Net      & 93.69$\pm$8.34  & 1.81$\pm$1.33 & 82.94$\pm$11.34 & 3.01$\pm$3.02 & 9.75  & 1.20 \\
Att U-Net  & 94.32$\pm$7.64  & 1.53$\pm$0.99 & 86.24$\pm$9.91  & 2.53$\pm$2.14 & 8.08  & 1.00 \\
nnU-Net    & 94.30$\pm$7.68  & 1.54$\pm$1.03 & 85.45$\pm$10.72 & 2.68$\pm$2.68 & 8.85  & 1.14 \\
nnFormer   & 93.95$\pm$7.52  & 1.95$\pm$3.12 & 84.15$\pm$10.38 & 3.17$\pm$2.52 & 9.80  & 1.22 \\
SegFormer  & 92.77$\pm$8.63  & 2.22$\pm$1.41 & 82.11$\pm$12.55 & 3.54$\pm$3.10 & 10.66 & 1.32 \\
TransBTS   & 94.05$\pm$8.11  & 1.60$\pm$1.14 & 84.87$\pm$11.09 & 2.70$\pm$2.74 & 9.18  & 1.10 \\
UNETR      & 92.54$\pm$10.22 & 2.27$\pm$1.91 & 80.65$\pm$14.01 & 3.85$\pm$3.22 & 11.89 & 1.58 \\
Swin UNETR & 93.80$\pm$8.57  & 1.70$\pm$1.31 & 83.82$\pm$12.20 & 2.76$\pm$2.99 & 9.98  & 1.06 \\
UX Net     & 93.92$\pm$8.43  & 1.69$\pm$1.51 & 82.90$\pm$11.51 & 3.07$\pm$2.95 & 11.02 & 1.38 \\
MedNeXt    & 94.19$\pm$7.82  & 1.61$\pm$1.18 & 85.92$\pm$10.88 & 2.67$\pm$2.70 & 8.27  & 1.06 \\
TransHRNet & 94.31$\pm$7.91  & 1.55$\pm$1.10 & 84.35$\pm$9.77  & 2.85$\pm$2.01 & 9.96  & 1.30 \\
VSmTrans   & 94.34$\pm$7.33  & 1.57$\pm$1.09 & 83.32$\pm$9.80  & 2.90$\pm$2.06 & 11.02 & 1.33 \\
MixUNETR   & 94.09$\pm$7.83  & 1.63$\pm$1.16 & 83.89$\pm$9.86  & 2.92$\pm$2.12 & 10.20 & 1.29 \\
\midrule
FEFormer   & \textbf{95.02}$^*\pm$5.96 & \textbf{1.40}$^*\pm$1.05 & \textbf{88.54}$^*\pm$8.72 & \textbf{2.32}$^*\pm$1.95 & 6.48 & 0.92 \\
\bottomrule
\end{tabular}}
\label{tab9}
\end{table}

\begin{figure*}[!t]
\centering
\includegraphics[width=\textwidth]{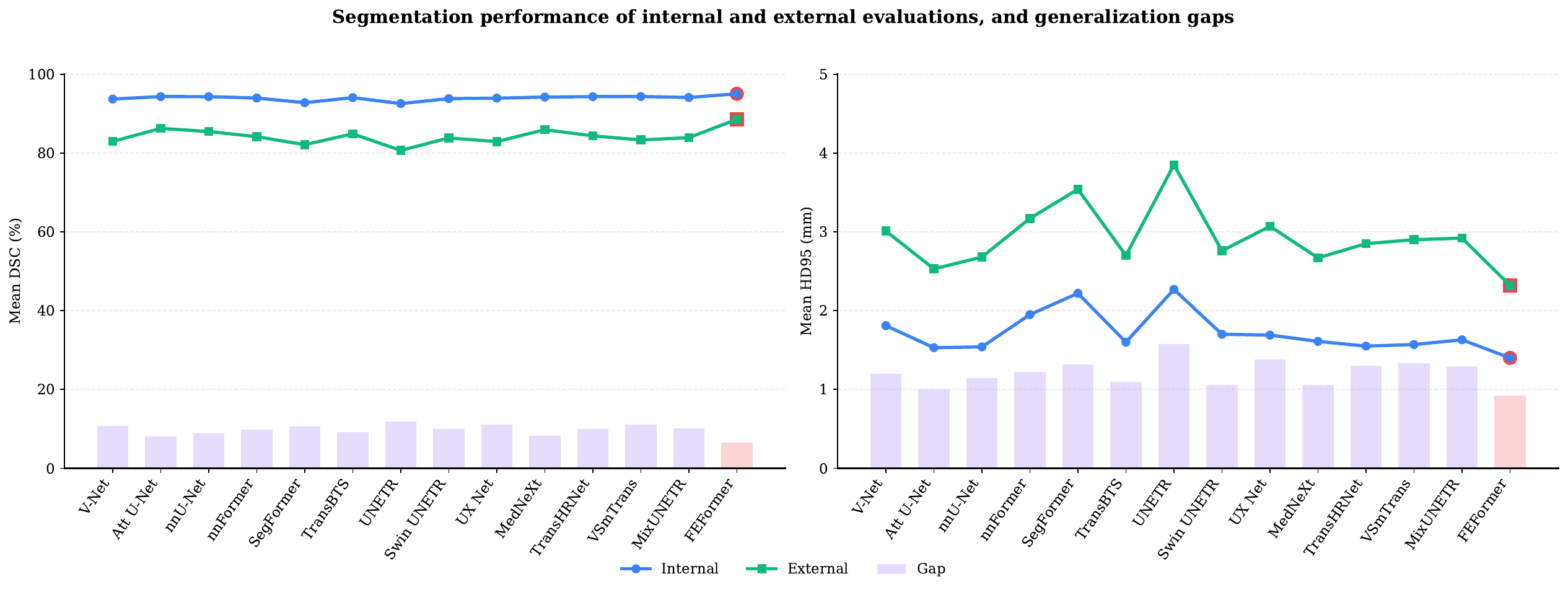}
\caption{Comparison of segmentation performance and generalization gaps between FEFormer and SOTA methods on the FLARE dataset for internal and external evaluation settings. Internal and external performance of all models are shown in DSC and HD95, along with the corresponding generalization gap. FEFormer consistently achieved superior segmentation accuracy (higher DSC and lower HD95) while exhibiting a notably smaller performance degradation between internal and external evaluation, indicating better robustness and generalization capability.}
\label{fig6}
\end{figure*}

\begin{figure*}[!t]
\centering
\includegraphics[width=\textwidth]{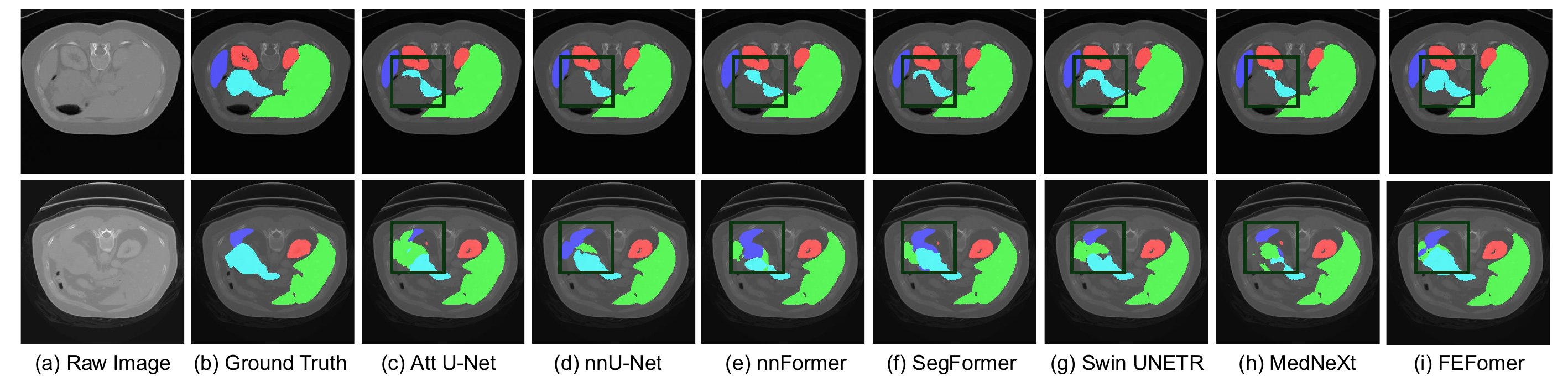}
\caption{Qualitative comparison between (i) FEFormer and (c) Att UNet, (d) nnU-Net, (e) nnFormer, (f) SegFormer, (g) Swin UNETR, and (h) MedNext on the external evaluation on the FLARE dataset. Red boxes mark the regions where FEFormer demonstrates better segmentation results than other methods.}
\label{fig7}
\end{figure*}

\subsection{Evaluation of generalization and robustness}
To further assess the robustness and cross-dataset generalizability of FEFormer, we conducted internal and external evaluations on the FLARE multi-organ segmentation task. In the internal evaluation, all methods were trained and tested within the FLARE dataset. In the external evaluation, models were trained on the AMOS dataset and directly applied to FLARE for zero-shot predictions without any fine-tuning, thus providing a more challenging assessment of domain generalization under distribution shift. The gaps of segmentation performance between internal and external evaluations were evaluated (Table~\ref{tab8} and Fig.~\ref{fig6}).

Specifically, FEFormer achieved the best performance on the internal and external evaluations and demonstrated the smallest generalization gap compared to other SOTA methods. Specifically, FEFormer achieved the highest mean DSC ($95.02\%$) and lowest mean HD95 ($1.40\%$) when the internal evaluation was performed to employ FEFormer within the FLARE dataset. Additionally, when FEFormer trained on AMOS were directly employed to FLARE for generating zero-shot predictions, FEFormer demonstrated substantially stronger robustness and generalizability to new domains than other methods. FEFormer achieved the highest mean DSC ($88.54\%$) and lowest mean HD95 ($2.32$ mm), outperforming CNN-based methods, including Att U-Net, nnU-Net and MedNeXt, in the internal evaluation. Additionally, FEFormer demonstrated higher generalizability than other hybrid models, such as TransHRNet, Swin UNETR, and MixUNETR. To quantify cross-domain robustness, we further measured the generalization gap, defined as the performance drop from internal to external evaluation. FEFormer exhibited the smallest DSC gap ($6.48$) and  DSC points and smallest HD95 increase ($0.92$), , whereas competing models showed markedly larger degradations (DSC gaps ranging from $8.08$ to $11.89$). This finding demonstrated that FEFormer maintained performance more consistently under domain shift and learned more generalizable representations by frequency modeling.

Qualitative comparisons among FEFormer and other methods further demonstrated its superior generalizability and robustness (Figure~\ref{fig7}). Specifically, when models were directly employed to the FLARE dataset for zero-shot predictions in the external evaluation, FEFormer exhibited less under-segmentation or miss-segmentation than comparing methods, and FEFormer demonstrated more accurate delineation of target organs by learning semantic features from other domains. Thus, FEFormer showed its superior anatomical consistency and robustness under external deployment conditions.

\begin{figure*}[!t]
\centering
\includegraphics[width=\textwidth]{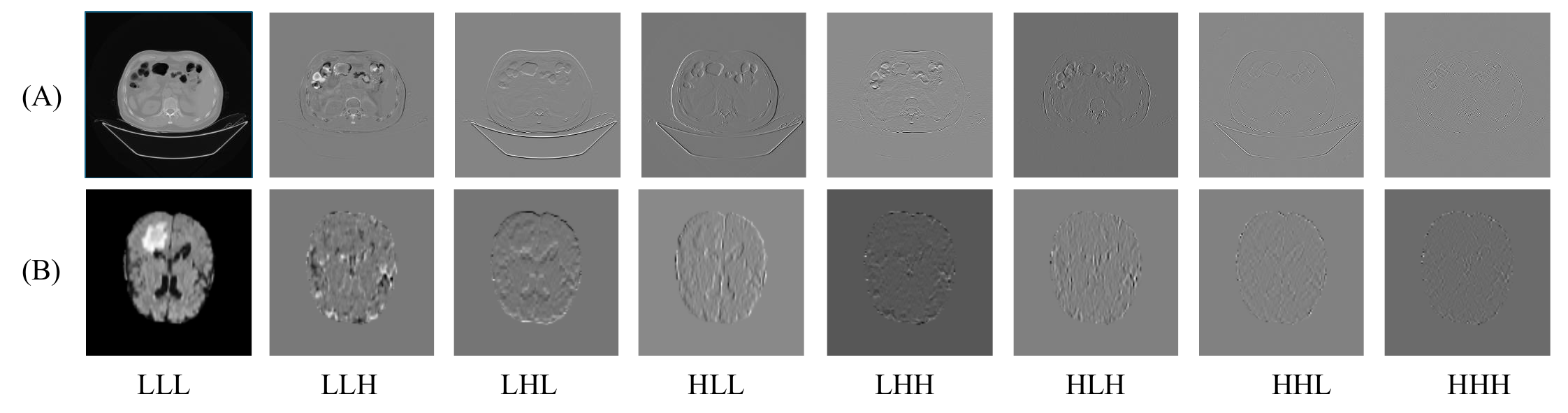}
\caption{Visualization of components decomposed by 3D Discrete Wavelet Transformation on (A) CT volumes from the AMOS dataset and (B) T1w MR volumes from the BraTS dataset.}
\label{fig8}
\end{figure*}

\subsection{Interpretability on architectural designs}
To systematically interpret the frequency characteristics in analyzing volumetric medical images, we visualized decompositions obtained from 3D DWT on CT volumes from the AMOS dataset and T1w MR volumes from the BraTS dataset (Figs.~\ref{fig8}). The 3D DWT provided a spatially localized, multi-resolution representation by decomposing each volume into one low-frequency component (LLL) and seven high-frequency sub-bands along three spatial directions (e.g, H, W, and D), thus preserving both spatial and frequency information. The LLL component preserved dominant anatomical structures and global intensity distributions, while high-frequency sub-bands encoded directional edge responses and fine-grained texture variations, highlighting boundaries and subtle anatomical details. Therefore, the WAFF module leveraged 3D DWT to decompose encoder and decoder features into semantically meaningful sub-bands, enabling frequency-aligned fusion that reduced large semantic discrepancies while preserving structural details. 

The 3D FFT converted spatial images and features to frequency-domain signals, and these frequency signals were decomposed into low-frequency and high-frequency components (Figs.~\ref{fig1}). Low-frequency components captured coarse structural information, while high-frequency components described peripheral regions and emphasized sharp intensity transitions and detailed patterns. Thus, the FDSA module performed global feature interaction in the frequency domain via FFT, enabling efficient long-range dependency modeling with $O(N\log N)$ complexity while preserving high-frequency signals critical for boundary delineation and detailed structure modeling. The FGMLP module disentangled spatial features into low- and high-frequency components and adaptively re-weighted them, enhancing discriminative feature learning across frequency bands. Additionally, the FCSB module integrated frequency-domain global interactions into early feature propagation, strengthening low-level feature consistency and improving fine structure recovery.

Motivated by these observations, FEFormer was designed to explicitly incorporate frequency-aware mechanisms at multiple levels, and these designs enabled FEFormer to effectively unify global spectral modeling and localized frequency decomposition, resulting in more robust and discriminative representations for volumetric medical image segmentation.

\section{Conclusion}
In this work, we proposed a Frequency-enhanced Vision Transformer for robust and efficient volumetric medical image segmentation, termed FEFormer. The Frequency-enhanced Dynamic Self-Attention (FDSA) and the Frequency-decomposed Gating MLP modules were developed and incorporated into the Frequency-enhanced Transformer block for generic feature extraction. Additionally, a Wavelet-guided Adaptive Feature Fusion (WAFF) module was proposed to fuse features by aligning their frequency components based on their semantics. A Frequency-enabled Cross-scale Stem Bridge was developed to enable low-level feature propagation. By incorporating these modules into a hierarchical transformer architecture, FEFormer achieved effective multi-scale representation learning and improved generalizability across diverse segmentation tasks. Extensive experiments demonstrated that FEFormer outperformed SOTA methods in segmentation accuracy while maintaining competitively high efficiency.

\section*{Declaration of competing interests}
The authors declare that they have no known competing financial interests or personal relationships that could have appeared to influence the work reported in this paper.

\section*{CRediT authorship contribution statement}
JY: conceptualization, methodology, formal analysis, writing the original draft, reviewing, and editing, visualization; XY: conceptualization, writing, reviewing, and editing; PQ: conceptualization, writing, reviewing, and editing.

\section*{Data availability}
The data used in this study are publicly available.



\bibliographystyle{cas-model2-names}

\bibliography{cas-refs}

@article{wang2021annotation,
  title={Annotation-efficient deep learning for automatic medical image segmentation},
  author={Wang, Shanshan and Li, Cheng and Wang, Rongpin and Liu, Zaiyi and Wang, Meiyun and Tan, Hongna and Wu, Yaping and Liu, Xinfeng and Sun, Hui and Yang, Rui and others},
  journal={Nature communications},
  volume={12},
  number={1},
  pages={5915},
  year={2021},
  publisher={Nature Publishing Group UK London}
}

@article{azad2024medical,
  title={Medical image segmentation review: The success of u-net},
  author={Azad, Reza and Aghdam, Ehsan Khodapanah and Rauland, Amelie and Jia, Yiwei and Avval, Atlas Haddadi and Bozorgpour, Afshin and Karimijafarbigloo, Sanaz and Cohen, Joseph Paul and Adeli, Ehsan and Merhof, Dorit},
  journal={IEEE Transactions on Pattern Analysis and Machine Intelligence},
  volume={46},
  number={12},
  pages={10076--10095},
  year={2024},
  publisher={IEEE}
}

@inproceedings{ronneberger2015u,
  title={U-net: Convolutional networks for biomedical image segmentation},
  author={Ronneberger, Olaf and Fischer, Philipp and Brox, Thomas},
  booktitle={International Conference on Medical image computing and computer-assisted intervention},
  pages={234--241},
  year={2015},
  organization={Springer}
}

@article{li2018h,
  title={H-DenseUNet: hybrid densely connected UNet for liver and tumor segmentation from CT volumes},
  author={Li, Xiaomeng and Chen, Hao and Qi, Xiaojuan and Dou, Qi and Fu, Chi-Wing and Heng, Pheng-Ann},
  journal={IEEE transactions on medical imaging},
  volume={37},
  number={12},
  pages={2663--2674},
  year={2018},
  publisher={IEEE}
}

@inproceedings{yang2025dynamic,
  title={Dynamic U-Net: adaptively calibrate features for abdominal multiorgan segmentation},
  author={Yang, Jin and Marcus, Daniel S and Sotiras, Aristeidis},
  booktitle={Medical Imaging 2025: Computer-Aided Diagnosis},
  volume={13407},
  pages={326--334},
  year={2025},
  organization={SPIE}
}

@inproceedings{yang2025d2,
  title={D2-mlp: dynamic decomposed mlp mixer for medical image segmentation},
  author={Yang, Jin and Yang, Jing and Yu, Xiaobing and Qiu, Peijie and Prajapat, Sunil},
  booktitle={ICASSP 2025-2025 IEEE International Conference on Acoustics, Speech and Signal Processing (ICASSP)},
  pages={1--5},
  year={2025},
  organization={IEEE}
}

@article{yang2025dmc,
  title={DMC-Net: Lightweight Dynamic Multi-scale and Multi-resolution convolution network for pancreas segmentation in CT images},
  author={Yang, Jin and Marcus, Daniel S and Sotiras, Aristeidis},
  journal={Biomedical Signal Processing and Control},
  volume={109},
  pages={107896},
  year={2025},
  publisher={Elsevier}
}

@article{yang2025translk,
  title={TransLK-Net: Entangling Transformer and Large Kernel for Progressive and Collaborative Feature Encoding and Decoding in Medical Image Segmentation},
  author={Yang, Jin and Marcus, Daniel S and Sotiras, Aristeidis},
  journal={arXiv preprint arXiv:2511.17873},
  year={2025}
}

@article{yang2026d,
  title={D-net: Dynamic large kernel with dynamic feature fusion for volumetric medical image segmentation},
  author={Yang, Jin and Qiu, Peijie and Zhang, Yichi and Marcus, Daniel S and Sotiras, Aristeidis},
  journal={Biomedical Signal Processing and Control},
  volume={113},
  pages={108837},
  year={2026},
  publisher={Elsevier}
}

@inproceedings{dosovitskiy2021an,
title={An Image is Worth 16x16 Words: Transformers for Image Recognition at Scale},
author={Alexey Dosovitskiy and Lucas Beyer and Alexander Kolesnikov and Dirk Weissenborn and Xiaohua Zhai and Thomas Unterthiner and Mostafa Dehghani and Matthias Minderer and Georg Heigold and Sylvain Gelly and Jakob Uszkoreit and Neil Houlsby},
booktitle={International Conference on Learning Representations},
year={2021},
url={https://openreview.net/forum?id=YicbFdNTTy}
}

@article{you2022class,
  title={Class-aware adversarial transformers for medical image segmentation},
  author={You, Chenyu and Zhao, Ruihan and Liu, Fenglin and Dong, Siyuan and Chinchali, Sandeep and Topcu, Ufuk and Staib, Lawrence and Duncan, James},
  journal={Advances in Neural Information Processing Systems},
  volume={35},
  pages={29582--29596},
  year={2022}
}

@inproceedings{yan2022after,
  title={After-unet: Axial fusion transformer unet for medical image segmentation},
  author={Yan, Xiangyi and Tang, Hao and Sun, Shanlin and Ma, Haoyu and Kong, Deying and Xie, Xiaohui},
  booktitle={Proceedings of the IEEE/CVF winter conference on applications of computer vision},
  pages={3971--3981},
  year={2022}
}

@article{chen2023cta,
  title={CTA-UNet: CNN-transformer architecture UNet for dental CBCT images segmentation},
  author={Chen, Zeyu and Chen, Senyang and Hu, Fengjun},
  journal={Physics in Medicine \& Biology},
  volume={68},
  number={17},
  pages={175042},
  year={2023},
  publisher={IOP Publishing}
}

@inproceedings{liu2022isegformer,
  title={iSegFormer: interactive segmentation via transformers with application to 3D knee MR images},
  author={Liu, Qin and Xu, Zhenlin and Jiao, Yining and Niethammer, Marc},
  booktitle={International Conference on Medical Image Computing and Computer-Assisted Intervention},
  pages={464--474},
  year={2022},
  organization={Springer}
}

@article{pecco2024optimizing,
  title={Optimizing performance of transformer-based models for fetal brain MR image segmentation},
  author={Pecco, Nicol{\`o} and Della Rosa, Pasquale Anthony and Canini, Matteo and Nocera, Gianluca and Scifo, Paola and Cavoretto, Paolo Ivo and Candiani, Massimo and Falini, Andrea and Castellano, Antonella and Baldoli, Cristina},
  journal={Radiology: Artificial Intelligence},
  volume={6},
  number={6},
  pages={e230229},
  year={2024},
  publisher={Radiological Society of North America}
}

@article{li2024swincross,
  title={SwinCross: Cross-modal Swin transformer for head-and-neck tumor segmentation in PET/CT images},
  author={Li, Gary Y and Chen, Junyu and Jang, Se-In and Gong, Kuang and Li, Quanzheng},
  journal={Medical physics},
  volume={51},
  number={3},
  pages={2096--2107},
  year={2024},
  publisher={Wiley Online Library}
}

@article{wu2022fat,
  title={FAT-Net: Feature adaptive transformers for automated skin lesion segmentation},
  author={Wu, Huisi and Chen, Shihuai and Chen, Guilian and Wang, Wei and Lei, Baiying and Wen, Zhenkun},
  journal={Medical image analysis},
  volume={76},
  pages={102327},
  year={2022},
  publisher={Elsevier}
}

@inproceedings{pan2023adaptive,
  title={Adaptive template transformer for mitochondria segmentation in electron microscopy images},
  author={Pan, Yuwen and Luo, Naisong and Sun, Rui and Meng, Meng and Zhang, Tianzhu and Xiong, Zhiwei and Zhang, Yongdong},
  booktitle={Proceedings of the IEEE/CVF International Conference on Computer Vision},
  pages={21474--21484},
  year={2023}
}

@inproceedings{ren2022shunted,
  title={Shunted self-attention via multi-scale token aggregation},
  author={Ren, Sucheng and Zhou, Daquan and He, Shengfeng and Feng, Jiashi and Wang, Xinchao},
  booktitle={Proceedings of the IEEE/CVF conference on computer vision and pattern recognition},
  pages={10853--10862},
  year={2022}
}

@article{xu2025s2aformer,
  title={S2AFormer: Strip Self-Attention for Efficient Vision Transformer},
  author={Xu, Guoan and Huang, Wenfeng and Jia, Wenjing and Li, Jiamao and Gao, Guangwei and Qi, Guo-Jun},
  journal={IEEE Transactions on Image Processing},
  volume={34},
  pages={8243--8256},
  year={2025},
  publisher={IEEE}
}

@article{yan2025multi,
  title={Multi-scale convolutional attention frequency-enhanced transformer network for medical image segmentation},
  author={Yan, Shun and Yang, Benquan and Chen, Aihua and Zhao, Xiaoming and Zhang, Shiqing},
  journal={Information Fusion},
  volume={119},
  pages={103019},
  year={2025},
  publisher={Elsevier}
}

@inproceedings{zhang2023pha,
  title={PHA: Patch-wise high-frequency augmentation for transformer-based person re-identification},
  author={Zhang, Guiwei and Zhang, Yongfei and Zhang, Tianyu and Li, Bo and Pu, Shiliang},
  booktitle={Proceedings of the IEEE/CVF conference on computer vision and pattern recognition},
  pages={14133--14142},
  year={2023}
}

@inproceedings{hatamizadeh2023global,
  title={Global context vision transformers},
  author={Hatamizadeh, Ali and Yin, Hongxu and Heinrich, Greg and Kautz, Jan and Molchanov, Pavlo},
  booktitle={International conference on machine learning},
  pages={12633--12646},
  year={2023},
  organization={PMLR}
}

@inproceedings{zheng2023lightweight,
  title={Lightweight vision transformer with spatial and channel enhanced self-attention},
  author={Zheng, Jiahao and Yang, Longqi and Li, Yiying and Yang, Ke and Wang, Zhiyuan and Zhou, Jun},
  booktitle={Proceedings of the IEEE/CVF international conference on computer vision},
  pages={1492--1496},
  year={2023}
}

@inproceedings{liu2021swin,
  title={Swin transformer: Hierarchical vision transformer using shifted windows},
  author={Liu, Ze and Lin, Yutong and Cao, Yue and Hu, Han and Wei, Yixuan and Zhang, Zheng and Lin, Stephen and Guo, Baining},
  booktitle={Proceedings of the IEEE/CVF international conference on computer vision},
  pages={10012--10022},
  year={2021}
}

@article{wu2025low,
  title={Low-resolution self-attention for semantic segmentation},
  author={Wu, Yu-Huan and Zhang, Shi-Chen and Liu, Yun and Zhang, Le and Zhan, Xin and Zhou, Daquan and Feng, Jiashi and Cheng, Ming-Ming and Zhen, Liangli},
  journal={IEEE Transactions on Pattern Analysis and Machine Intelligence},
  year={2025},
  publisher={IEEE}
}

@article{song2026rmt,
  title={SU-RMT: Toward Bridging Semantic Representation and Structural Detail Modeling for Medical Image Segmentation},
  author={Song, Peibo and Wang, Zihao and Zhang, Jinshuo and Fu, Shujun and Zhang, Yunfeng and Wu, Wei and Bao, Fangxun},
  journal={Information Fusion},
  pages={104182},
  year={2026},
  publisher={Elsevier}
}

@inproceedings{bai2022improving,
  title={Improving vision transformers by revisiting high-frequency components},
  author={Bai, Jiawang and Yuan, Li and Xia, Shu-Tao and Yan, Shuicheng and Li, Zhifeng and Liu, Wei},
  booktitle={European conference on computer vision},
  pages={1--18},
  year={2022},
  organization={Springer}
}

@inproceedings{hu2018squeeze,
  title={Squeeze-and-excitation networks},
  author={Hu, Jie and Shen, Li and Sun, Gang},
  booktitle={Proceedings of the IEEE conference on computer vision and pattern recognition},
  pages={7132--7141},
  year={2018}
}

@inproceedings{wang2020eca,
  title={ECA-Net: Efficient channel attention for deep convolutional neural networks},
  author={Wang, Qilong and Wu, Banggu and Zhu, Pengfei and Li, Peihua and Zuo, Wangmeng and Hu, Qinghua},
  booktitle={Proceedings of the IEEE/CVF conference on computer vision and pattern recognition},
  pages={11534--11542},
  year={2020}
}

@inproceedings{zhong2020squeeze,
  title={Squeeze-and-attention networks for semantic segmentation},
  author={Zhong, Zilong and Lin, Zhong Qiu and Bidart, Rene and Hu, Xiaodan and Daya, Ibrahim Ben and Li, Zhifeng and Zheng, Wei-Shi and Li, Jonathan and Wong, Alexander},
  booktitle={Proceedings of the IEEE/CVF conference on computer vision and pattern recognition},
  pages={13065--13074},
  year={2020}
}

@inproceedings{cciccek20163d,
  title={3D U-Net: learning dense volumetric segmentation from sparse annotation},
  author={{\c{C}}i{\c{c}}ek, {\"O}zg{\"u}n and Abdulkadir, Ahmed and Lienkamp, Soeren S and Brox, Thomas and Ronneberger, Olaf},
  booktitle={International conference on medical image computing and computer-assisted intervention},
  pages={424--432},
  year={2016},
  organization={Springer}
}

@article{sun2025msm,
  title={MSM-UNet: a medical image segmentation method based on wavelet transform and multi-scale Mamba-UNet},
  author={Sun, Junding and Chen, Kaixin and Wu, Xiaosheng and Xu, Zhaozhao and Wang, Shuihua and Zhang, Yudong},
  journal={Expert Systems with Applications},
  volume={288},
  pages={128241},
  year={2025},
  publisher={Elsevier}
}

@article{xu2025x,
  title={X-UNet: A novel global context-aware collaborative fusion U-shaped network with progressive feature fusion of codec for medical image segmentation},
  author={Xu, Shijie and Chen, Yufeng and Zhang, Xiaoqian and Sun, Feng and Chen, Siyu and Ou, Yanchi and Luo, Chao},
  journal={Neural Networks},
  pages={107943},
  year={2025},
  publisher={Elsevier}
}

@article{xiao2021early,
  title={Early convolutions help transformers see better},
  author={Xiao, Tete and Singh, Mannat and Mintun, Eric and Darrell, Trevor and Doll{\'a}r, Piotr and Girshick, Ross},
  journal={Advances in neural information processing systems},
  volume={34},
  pages={30392--30400},
  year={2021}
}

@article{chi2020fast,
  title={Fast fourier convolution},
  author={Chi, Lu and Jiang, Borui and Mu, Yadong},
  journal={Advances in Neural Information Processing Systems},
  volume={33},
  pages={4479--4488},
  year={2020}
}

@inproceedings{wang2022mixed,
  title={Mixed transformer u-net for medical image segmentation},
  author={Wang, Hongyi and Xie, Shiao and Lin, Lanfen and Iwamoto, Yutaro and Han, Xian-Hua and Chen, Yen-Wei and Tong, Ruofeng},
  booktitle={ICASSP 2022-2022 IEEE international conference on acoustics, speech and signal processing (ICASSP)},
  pages={2390--2394},
  year={2022},
  organization={IEEE}
}

@article{zhang2025lightweight,
  title={A lightweight convolution and vision transformer integrated model with multi-scale self-attention mechanism},
  author={Zhang, Yi and Wei, Lingxiao and Zhang, Bowei and Liu, Ziwei and Yi, Kai and Hu, Shu},
  journal={Neurocomputing},
  pages={131670},
  year={2025},
  publisher={Elsevier}
}

@article{xiao2023transformers,
  title={Transformers in medical image segmentation: A review},
  author={Xiao, Hanguang and Li, Li and Liu, Qiyuan and Zhu, Xiuhong and Zhang, Qihang},
  journal={Biomedical Signal Processing and Control},
  volume={84},
  pages={104791},
  year={2023},
  publisher={Elsevier}
}

@inproceedings{cao2022swin,
  title={Swin-unet: Unet-like pure transformer for medical image segmentation},
  author={Cao, Hu and Wang, Yueyue and Chen, Joy and Jiang, Dongsheng and Zhang, Xiaopeng and Tian, Qi and Wang, Manning},
  booktitle={European conference on computer vision},
  pages={205--218},
  year={2022},
  organization={Springer}
}

@article{huang2022missformer,
  title={Missformer: An effective transformer for 2d medical image segmentation},
  author={Huang, Xiaohong and Deng, Zhifang and Li, Dandan and Yuan, Xueguang and Fu, Ying},
  journal={IEEE Transactions on Medical Imaging},
  volume={42},
  number={5},
  pages={1484--1494},
  year={2022},
  publisher={IEEE}
}

@article{lin2022ds,
  title={Ds-transunet: Dual swin transformer u-net for medical image segmentation},
  author={Lin, Ailiang and Chen, Bingzhi and Xu, Jiayu and Zhang, Zheng and Lu, Guangming and Zhang, David},
  journal={IEEE Transactions on Instrumentation and Measurement},
  volume={71},
  pages={1--15},
  year={2022},
  publisher={IEEE}
}

@inproceedings{valanarasu2021medical,
  title={Medical transformer: Gated axial-attention for medical image segmentation},
  author={Valanarasu, Jeya Maria Jose and Oza, Poojan and Hacihaliloglu, Ilker and Patel, Vishal M},
  booktitle={Medical image computing and computer assisted intervention--MICCAI 2021: 24th international conference, Strasbourg, France, September 27--October 1, 2021, proceedings, part I 24},
  pages={36--46},
  year={2021},
  organization={Springer}
}

@article{zhang2024ct,
  title={Ct-net: Asymmetric compound branch transformer for medical image segmentation},
  author={Zhang, Ning and Yu, Long and Zhang, Dezhi and Wu, Weidong and Tian, Shengwei and Kang, Xiaojing and Li, Min},
  journal={Neural Networks},
  volume={170},
  pages={298--311},
  year={2024},
  publisher={Elsevier}
}

@article{li2025seaformer,
  title={SEAformer: Selective Edge Aggregation transformer for 2D medical image segmentation},
  author={Li, Jingwen and Chen, Jilong and Li, Ruoyu and Han, Peilun and Cheng, Junlong and others},
  journal={Biomedical Signal Processing and Control},
  volume={102},
  pages={107203},
  year={2025},
  publisher={Elsevier}
}

@article{qiu2026agileformer,
  title={AgileFormer: Spatially agile and scalable transformer for medical image segmentation},
  author={Qiu, Peijie and Yang, Jin and Kumar, Sayantan and Ghosh, Soumyendu Sekhar and Sotiras, Aristeidis},
  journal={Biomedical Signal Processing and Control},
  volume={112},
  pages={108842},
  year={2026},
  publisher={Elsevier}
}

@inproceedings{patro2025spectformer,
  title={Spectformer: Frequency and attention is what you need in a vision transformer},
  author={Patro, Badri N and Namboodiri, Vinay P and Agneeswaran, Vijay S},
  booktitle={2025 IEEE/CVF winter conference on applications of computer vision (WACV)},
  pages={9543--9554},
  year={2025},
  organization={IEEE}
}

@inproceedings{dai2024freqformer,
  title={FreqFormer: Frequency-aware Transformer for Lightweight Image Super-resolution.},
  author={Dai, Tao and Wang, Jianping and Guo, Hang and Li, Jinmin and Wang, Jinbao and Zhu, Zexuan},
  booktitle={IJCAI},
  pages={731--739},
  year={2024}
}

@article{song2025efficient,
  title={Efficient frequency feature aggregation transformer for image super-resolution},
  author={Song, Jianwen and Sowmya, Arcot and Sun, Changming},
  journal={Pattern Recognition},
  volume={167},
  pages={111735},
  year={2025},
  publisher={Elsevier}
}

@article{shang2024holistic,
  title={Holistic dynamic frequency transformer for image fusion and exposure correction},
  author={Shang, Xiaoke and Li, Gehui and Jiang, Zhiying and Zhang, Shaomin and Ding, Nai and Liu, Jinyuan},
  journal={Information Fusion},
  volume={102},
  pages={102073},
  year={2024},
  publisher={Elsevier}
}

@inproceedings{mao2024loformer,
  title={Loformer: Local frequency transformer for image deblurring},
  author={Mao, Xintian and Wang, Jiansheng and Xie, Xingran and Li, Qingli and Wang, Yan},
  booktitle={Proceedings of the 32nd ACM International Conference on Multimedia},
  pages={10382--10391},
  year={2024}
}

@inproceedings{kong2023efficient,
  title={Efficient frequency domain-based transformers for high-quality image deblurring},
  author={Kong, Lingshun and Dong, Jiangxin and Ge, Jianjun and Li, Mingqiang and Pan, Jinshan},
  booktitle={Proceedings of the IEEE/CVF conference on computer vision and pattern recognition},
  pages={5886--5895},
  year={2023}
}

@article{zeng2024dbfft,
  title={DBFFT: Adversarial-robust dual-branch frequency domain feature fusion in vision transformers},
  author={Zeng, Jia and Huang, Lan and Bai, Xingyu and Wang, Kangping},
  journal={Information Fusion},
  volume={108},
  pages={102387},
  year={2024},
  publisher={Elsevier}
}

@inproceedings{pan2022wnet,
  title={Wnet: Audio-guided video object segmentation via wavelet-based cross-modal denoising networks},
  author={Pan, Wenwen and Shi, Haonan and Zhao, Zhou and Zhu, Jieming and He, Xiuqiang and Pan, Zhigeng and Gao, Lianli and Yu, Jun and Wu, Fei and Tian, Qi},
  booktitle={Proceedings of the IEEE/CVF conference on computer vision and pattern recognition},
  pages={1320--1331},
  year={2022}
}

@inproceedings{zhou2023xnet,
  title={Xnet: Wavelet-based low and high frequency fusion networks for fully-and semi-supervised semantic segmentation of biomedical images},
  author={Zhou, Yanfeng and Huang, Jiaxing and Wang, Chenlong and Song, Le and Yang, Ge},
  booktitle={Proceedings of the IEEE/CVF international conference on computer vision},
  pages={21085--21096},
  year={2023}
}

@article{zhan2025wcmamba,
  title={WCMamba: Enhancing high-resolution remote sensing image semantic segmentation with pyramid wavelet convolution and SS2D},
  author={Zhan, Chao and Yang, Kui},
  journal={Knowledge-Based Systems},
  volume={324},
  pages={113877},
  year={2025},
  publisher={Elsevier}
}

@article{xiao2026wtclip,
  title={WTCLIP: A Wavelet-Aware CLIP Framework for Boundary-Refined Weakly Supervised Semantic Segmentation},
  author={Xiao, Feng and Zhang, Jianhua and Han, Peihua and Chen, Shengyong and Zhang, Houxiang},
  journal={IEEE Transactions on Industrial Informatics},
  year={2026},
  publisher={IEEE}
}

@article{li2026frequency,
  title={Frequency Domain-Enhanced Spectral-Spatial Fusion Transformer for Semantic Segmentation of Remote Sensing Images},
  author={Li, Xin and Xu, Feng and Li, Jiaxin and Su, Yuanchao and Li, Linyi and Lyu, Xin and Xu, Zhennan and Kaup, Andr{\'e}},
  journal={Information Fusion},
  pages={104248},
  year={2026},
  publisher={Elsevier}
}

@article{ji2022amos,
  title={Amos: A large-scale abdominal multi-organ benchmark for versatile medical image segmentation},
  author={Ji, Yuanfeng and Bai, Haotian and Ge, Chongjian and Yang, Jie and Zhu, Ye and Zhang, Ruimao and Li, Zhen and Zhanng, Lingyan and Ma, Wanling and Wan, Xiang and others},
  journal={Advances in neural information processing systems},
  volume={35},
  pages={36722--36732},
  year={2022}
}

@article{antonelli2022medical,
  title={The medical segmentation decathlon},
  author={Antonelli, Michela and Reinke, Annika and Bakas, Spyridon and Farahani, Keyvan and Kopp-Schneider, Annette and Landman, Bennett A and Litjens, Geert and Menze, Bjoern and Ronneberger, Olaf and Summers, Ronald M and others},
  journal={Nature communications},
  volume={13},
  number={1},
  pages={4128},
  year={2022},
  publisher={Nature Publishing Group UK London}
}

@article{ma2022fast,
  title={Fast and low-GPU-memory abdomen CT organ segmentation: the flare challenge},
  author={Ma, Jun and Zhang, Yao and Gu, Song and An, Xingle and Wang, Zhihe and Ge, Cheng and Wang, Congcong and Zhang, Fan and Wang, Yu and Xu, Yinan and others},
  journal={Medical Image Analysis},
  volume={82},
  pages={102616},
  year={2022},
  publisher={Elsevier}
}

@inproceedings{milletari2016v,
  title={V-net: Fully convolutional neural networks for volumetric medical image segmentation},
  author={Milletari, Fausto and Navab, Nassir and Ahmadi, Seyed-Ahmad},
  booktitle={2016 fourth international conference on 3D vision (3DV)},
  pages={565--571},
  year={2016},
  organization={Ieee}
}

@article{isensee2021nnu,
  title={nnU-Net: a self-configuring method for deep learning-based biomedical image segmentation},
  author={Isensee, Fabian and Jaeger, Paul F and Kohl, Simon AA and Petersen, Jens and Maier-Hein, Klaus H},
  journal={Nature methods},
  volume={18},
  number={2},
  pages={203--211},
  year={2021},
  publisher={Nature Publishing Group}
}

@article{oktay2018attention,
  title={Attention u-net: Learning where to look for the pancreas},
  author={Oktay, Ozan and Schlemper, Jo and Folgoc, Loic Le and Lee, Matthew and Heinrich, Mattias and Misawa, Kazunari and Mori, Kensaku and McDonagh, Steven and Hammerla, Nils Y and Kainz, Bernhard and others},
  journal={arXiv preprint arXiv:1804.03999},
  year={2018}
}

@inproceedings{wang2021transbts,
  title={Transbts: Multimodal brain tumor segmentation using transformer},
  author={Wang, Wenxuan and Chen, Chen and Ding, Meng and Yu, Hong and Zha, Sen and Li, Jiangyun},
  booktitle={Medical Image Computing and Computer Assisted Intervention--MICCAI 2021: 24th International Conference, Strasbourg, France, September 27--October 1, 2021, Proceedings, Part I 24},
  pages={109--119},
  year={2021},
  organization={Springer}
}

@inproceedings{hatamizadeh2022unetr,
  title={Unetr: Transformers for 3d medical image segmentation},
  author={Hatamizadeh, Ali and Tang, Yucheng and Nath, Vishwesh and Yang, Dong and Myronenko, Andriy and Landman, Bennett and Roth, Holger R and Xu, Daguang},
  booktitle={Proceedings of the IEEE/CVF winter conference on applications of computer vision},
  pages={574--584},
  year={2022}
}

@inproceedings{hatamizadeh2021swin,
  title={Swin unetr: Swin transformers for semantic segmentation of brain tumors in mri images},
  author={Hatamizadeh, Ali and Nath, Vishwesh and Tang, Yucheng and Yang, Dong and Roth, Holger R and Xu, Daguang},
  booktitle={International MICCAI Brainlesion Workshop},
  pages={272--284},
  year={2021},
  organization={Springer}
}

@article{lee20223d,
  title={3d ux-net: A large kernel volumetric convnet modernizing hierarchical transformer for medical image segmentation},
  author={Lee, Ho Hin and Bao, Shunxing and Huo, Yuankai and Landman, Bennett A},
  journal={arXiv preprint arXiv:2209.15076},
  year={2022}
}

@inproceedings{roy2023mednext,
  title={Mednext: transformer-driven scaling of convnets for medical image segmentation},
  author={Roy, Saikat and Koehler, Gregor and Ulrich, Constantin and Baumgartner, Michael and Petersen, Jens and Isensee, Fabian and Jaeger, Paul F and Maier-Hein, Klaus H},
  booktitle={International Conference on Medical Image Computing and Computer-Assisted Intervention},
  pages={405--415},
  year={2023},
  organization={Springer}
}

@inproceedings{perera2024segformer3d,
  title={SegFormer3D: an Efficient Transformer for 3D Medical Image Segmentation},
  author={Perera, Shehan and Navard, Pouyan and Yilmaz, Alper},
  booktitle={Proceedings of the IEEE/CVF Conference on Computer Vision and Pattern Recognition},
  pages={4981--4988},
  year={2024}
}

@article{zhou2023nnformer,
  title={nnformer: Volumetric medical image segmentation via a 3d transformer},
  author={Zhou, Hong-Yu and Guo, Jiansen and Zhang, Yinghao and Han, Xiaoguang and Yu, Lequan and Wang, Liansheng and Yu, Yizhou},
  journal={IEEE Transactions on Image Processing},
  year={2023},
  publisher={IEEE}
}

@article{liu2024vsmtrans,
  title={VSmTrans: A Hybrid Paradigm Integrating Self-attention and Convolution for 3D Medical Image Segmentation},
  author={Liu, Tiange and Bai, Qingze and Torigian, Drew A and Tong, Yubing and Udupa, Jayaram K},
  journal={Medical Image Analysis},
  pages={103295},
  year={2024},
  publisher={Elsevier}
}

@article{shen2025mixunetr,
  title={MixUNETR: A U-shaped network based on W-MSA and depth-wise convolution with channel and spatial interactions for zonal prostate segmentation in MRI},
  author={Shen, Quanyou and Zheng, Bowen and Li, Wenhao and Shi, Xiaoran and Luo, Kun and Yao, Yuqian and Li, Xinyan and Lv, Shidong and Tao, Jie and Wei, Qiang},
  journal={Neural Networks},
  volume={181},
  pages={106782},
  year={2025},
  publisher={Elsevier}
}

@article{yan20233d,
  title={3D medical image segmentation using parallel transformers},
  author={Yan, Qingsen and Liu, Shengqiang and Xu, Songhua and Dong, Caixia and Li, Zongfang and Shi, Javen Qinfeng and Zhang, Yanning and Dai, Duwei},
  journal={Pattern Recognition},
  volume={138},
  pages={109432},
  year={2023},
  publisher={Elsevier}
}

@inproceedings{dai2021attentional,
  title={Attentional feature fusion},
  author={Dai, Yimian and Gieseke, Fabian and Oehmcke, Stefan and Wu, Yiquan and Barnard, Kobus},
  booktitle={Proceedings of the IEEE/CVF winter conference on applications of computer vision},
  pages={3560--3569},
  year={2021}
}



\end{document}